\documentclass[10pt, onecolumn]{article}

\usepackage{amsmath}
\usepackage{amssymb}

\usepackage{graphicx}

\usepackage{cite}

\usepackage{color} 

\usepackage[labelfont=bf,labelsep=period,justification=raggedright]{caption}

\makeatletter
\renewcommand{\@biblabel}[1]{\quad#1.}
\makeatother

\date{}

\newcommand{\heaviside}{\mathop{\mathrm{\vartheta}}\nolimits}
\begin{document}
\begin{flushleft}
{\Large
\textbf{The fate of cooperation during range expansions}
}
\\
Kirill S. Korolev$^{1,\ast}$
\\
\bf{1} Massachusetts Institute of Technology, Department of Physics, Cambridge, MA 02139
\\
$\ast$ E-mail: papers.korolev@gmail.com
\end{flushleft}

\section*{Abstract}

Species expand their geographical ranges following an environmental change, long range dispersal, or a new adaptation. Range expansions not only bring an ecological change, but also affect the evolution of the expanding species. Although the dynamics of deleterious, neutral, and beneficial mutations have been extensively studied in expanding populations, the fate of alleles under frequency-dependent selection remains largely unexplored. The dynamics of cooperative alleles are particularly interesting because selection can be both frequency and density dependent resulting in a coupling between population and evolutionary dynamics. This coupling leads to an increase in the frequency of cooperators at the expansion front, and, under certain conditions, the entire front can be taken over by cooperators. Thus, a mixed population wave can split into an expansion wave of only cooperators followed by an invasion wave of defectors. After the splitting, cooperators increase in abundance by expanding into new territories faster than they are invaded by defectors. Our results not only provide an explanation for the maintenance of cooperation, but also elucidate the effect of eco-evolutionary feedback on the maintenance of genetic diversity during range expansions. When cooperators do not split away, we find that defectors can spread much faster with cooperators than they would be able to on their own or by invading cooperators. This enhanced rate of expansion in mixed waves could counterbalance the loss of genetic diversity due to the founder effect for mutations under frequency-dependent selection. Although we focus on cooperator-defector interactions, our analysis could also be relevant for other systems described by reaction-diffusion equations.    

\section*{Author Summary}

Cooperation is beneficial for the species as a whole, but, at the level of an individual, defection pays off. Natural selection is then expected to favor defectors and eliminate cooperation. This prediction is in stark contrast with the abundance of cooperation at all levels of biological systems: from bacterial biofilms to ecosystems and human societies. Several explanations have been proposed to resolve this paradox, including direct reciprocity and group selection. Our work, however, builds upon an observation that natural selection on cooperators might depend both on their relative frequency in the population and on the population density. We find that this feedback between the population and evolutionary dynamics can substantially increase the frequency of cooperators at the front of an expanding population, and can even lead to a splitting of cooperators from defectors. After splitting, only cooperators colonize new territories, while defectors slowly invade them from behind. Since range expansions are very common in nature, our work provides a new explanation of the maintenance of cooperation. More generally, the phenomena we describe could be of interest in other situations when coexisting entities spread in space, be it species in ecology or diffusing and reacting molecules in chemical kinetics.      

\section*{Introduction}

Cooperation between organisms has always interested evolutionary biologists~\cite{gardner:cooperation_history,nowak:book,smith:games}. On the one hand, cooperative interactions are widespread in living systems. Microbes cooperate to digest food, scavenge for scarce resources, or build a protective biofilm~\cite{west:social}; animals cooperate to hunt or avoid predation~\cite{holldobler:ants}; and individual cells cooperate to enable multicellular life~\cite{michod:multicellularity, sachs:multicellularity}. On the other hand, evolutionary theory has struggled to explain the mere existence of cooperation~\cite{gardner:cooperation_history,nowak:book,smith:games}. Although cooperation is beneficial to the species, it is susceptible to invasion by defectors, individuals who reap the benefits of cooperation without paying the costs. Defectors, having a higher relative fitness, are expected to take over the population leading to the demise of cooperation. The breakdown of cooperation often has devastating consequences such as the tragedy of the commons~\cite{hardin:tragedy} or cancer~\cite{crespi:cancer_evolution, merlo:cancer_evolution}. The break down of cooperation can also be desirable, e.g., when it destroys biofilms protecting pathogens from antibiotics and the immune system~\cite{west:social}. Understanding the evolution and maintenance of cooperation is therefore an important problem in economics, medicine, and biology.  

Several mechanisms have been proposed that can stabilize cooperation against defection~\cite{gardner:cooperation_history,nowak:book,nowak:five_rules}, including direct reciprocity~\cite{axelrod:cooperation,nowak:generous_tft}, group and kin selection~\cite{wilson:group_selection,hamilton:rule1,hamilton:rule2}, and spatial structure~\cite{nowak:spatial,roca:spatial_cooperation,nadell:EmergenceSpatialStructure}. Most of the previous studies have focused exclusively on the changes in the relative frequencies of cooperators and defectors and neglected possible changes in the population size. The naive expectation, however, is that a reduced level of cooperation should lead to a lower average fitness of the population and, therefore, to a lower population size. Indeed, this effect of evolutionary on ecological dynamics has been observed in several experimental populations~\cite{turner:rna, rainey:surface_film, griffin:pathogenic} as well as in ecological public goods games~\cite{hauert:prs_public_goods,hauert:tpb_public_goods,wakano:spatial_public_goods}. In the latter studies, the authors also found that, under certain conditions, cooperators are favored by natural selection at low population densities while defectors are favored at high population densities. This dependence of evolutionary dynamics on population density suggests that the low-density edges of population ranges might be conducive to the evolution of cooperation. The edge of an expanding population is of particular interest because the new territories might be colonized mostly by cooperators.

Range expansions and range shifts are common in nature~\cite{shine:spatial_sorting, templeton:africa, parmesan:butterfly_climate, pateman:butterfly_climate, gray:worm_expansion, phillips:toad_acceleration,hitch:climate_birds, bronnenhuber:goby_invasion}. Examples include the recolonizations of temperate latitudes between glaciations, the invasion of North American forests by the Asian long-horned beetle~(\textit{Anoplophora glabripennis}), and the spread of the western corn rootworm~(\textit{Diabrotica virgifera}) in the Midwestern United States. Apart from ecological and sometimes economic impact, range expansions bring about significant changes in the genetic diversity and the evolutionary dynamics of the expanding species~\cite{excoffier:review, korolev:review, vlad:hydrodynamics, korolev:mutualism, mcinerny:climate_shift, roques:allee_diversity, arenas:diversity}. In particular, genetic diversity is typically lost because of the founder effect, and mutations appearing close to the expansion edge are very likely to reach high frequencies in the population even if they are neutral or slightly deleterious~\cite{excoffier:review,hallatschek:1dwave,hallatschek:life_front,roques:allee_diversity,travis:deleterious_surfing}. Although the dynamics of neutral, deleterious, and beneficial mutations arising at the edge of a range expansion have been extensively studied, the fate of mutations subject to frequency-dependent selection, e.g. encoding cooperative traits, has received little attention.  

Here, we study how the interplay between ecological and evolutionary processes affects the evolution of cooperation during range expansions. To this end, we formulate a model that combines the effect of genetic composition on population growth and the effect of population density on the fitnesses of different genotypes. We find that cooperators are favored at the edge of an expanding population and, under certain conditions, cooperators can outrun defectors and spread into unoccupied territories, leaving defectors behind. This mechanism of maintaining cooperation might play an important role in populations that experience frequent disturbances~(e.g. forest fires) followed by range expansions. More generally, we report the splitting of a mixed two-species~(or two-allele) traveling wave into a wave of one species followed by an invasion wave of the other species. This wave-splitting phenomenon might also be relevant to other processes described by traveling reaction-diffusion waves, which are widely used in biology and chemical kinetics~\cite{murray:mathematical_biology,vlad:hydrodynamics,douglas:assembly_wave}.

Our works complements two previous studies of the evolutionary dynamics of cooperation in spatial populations~\cite{wakano:spatial_public_goods, sella:dynamic_persistence}. In~\cite{wakano:spatial_public_goods}, the intricate spatial patters of coexisting cooperators and defectors were reported  for ecological public goods games. In~\cite{sella:dynamic_persistence}, the authors found that cooperation can persist in a spatial prisoner's dilemma game because of the cyclic turnover of cooperators, defectors, and extinctions. Similar to our work, their results rely on the fact that cooperators grow faster than defectors when spatially isolated from each other.

When splitting does not occur, defectors slow down cooperators, and the rate of spreading is primarily controlled by the evolutionary dynamics at the front. For example, we find that a mixed wave of cooperators and defectors can experience long periods of acceleration, as the genetic composition at the front adjusts to the low-density conditions. We also find that defectors can spread faster in mixed waves than they would be able in isolation or by invading cooperators. This finding could be important for conservation efforts to help ecosystems shift cohesively in response to a rapid climate change or habitat deterioration.

\section*{Models}

To understand the fate of cooperation during range expansions, we need a spatial model of a mixed population of cooperators and defectors colonizing new territories. We first discuss range expansions of pure cooperators and then consider defectors invading cooperators. Although these two spreading phenomena have a similar mathematical description, they have rarely been studied together, because one is an ecological, and the other is an evolutionary process. Finally, we introduce a complete model that allows for changes in both population density and allele frequencies and explicitly includes the coupling between these two variables. For simplicity, only one-dimensional expansion waves are considered, which should be a good approximation when number fluctuations and the curvature of the wave front can be neglected. The theoretical studies of range expansions were pioneered in~\cite{fisher:wave, kolmogorov:wave}, and a good introduction to this topic can be found in~\cite{murray:mathematical_biology}.

\subsection*{Ecological dynamics}

We begin by considering populations of only cooperators. Range expansions are driven by migration from colonized to new territories and by population growth. When migration~(or dispersal) is short-range and isotropic, it can be approximated by a diffusion term leading to the following reaction-diffusion model of range expansions

\begin{equation}
	\frac{\partial c}{\partial t} = D\frac{\partial^{2} c}{\partial x^{2}} + G_{c}(c)c,
	\label{eq:dcdt_sp}
\end{equation}

\noindent where~$c(t,x)$ is the population density at time~$t$ and spatial location~$x$,~$D$~is the effective diffusion constant, and~$G_{c}(c)$ is the per capita growth rate.

Since any habitat has a limited carrying capacity, the per capita growth rate, $G_{c}(c)$, must decline and become negative at high population densities. Populations with cooperative growth may also experience reduced or negative growth rates at low population densities because, for small~$c$, the probability of forming cooperating groups or the size of these groups is too small. Such nonmonotonic dependence of the per capita growth rate on population density is called an Allee effect and has been observed in different species ranging from budding yeast to desert bighorn sheep~\cite{allee:effect, courchamp:allee_review, berec:multiple_allee, dai:warnings, courchamp:dog, cluttonbrock:mongoose, mooring:sheep}. The most common model of an Allee effect assumes the following growth rate~\cite{murray:mathematical_biology,courchamp:allee_review} 

\begin{equation}
	G_{c}(c)c = g_{c}c(K-c)(c-c^{*}),
	\label{eq:dcdt_wm}
\end{equation}

\noindent where~$K$ is the carrying capacity, $c^{*}$ is the Allee threshold, i.e. the minimal density required for populations to grow, and~$g_{c}K^{2}$ sets the overall magnitude of the per capita growth rate. One typically distinguishes a strong Allee effect when~$c^{*}>0$ and a weak Allee effect when~$-K<c^{*}<0$. Allee effect is absent when~$c^{*}<-K$ because~$G_{c}(c)$ monotonically decreases for~$c>0$. Thus, equation~(\ref{eq:dcdt_wm}) is sufficiently flexible to describe populations with and without an Allee effect.

Even if the growth rate is negative at small densities, populations can spread into unoccupied territories. At~$t=0$, we assume that the habitat is colonized for all~$x<0$, but it is empty for~$x>0$. After an initial transitory period, expansion waves typically move at a constant velocity and the density profile does not change in the comoving reference frame, i.e. the reference frame moving along with the expansion wave; see figure~\ref{fig:three_velocities}ab. The expansion velocity is known exactly~\cite{murray:mathematical_biology,kolmogorov:wave,fisher:wave,fife:allee_wave,aronson:allee_wave} and is given by

\begin{equation}
	v_{\rm{c}}=
\left\{
\begin{array}{ll}
	 \sqrt{\frac{Dg_{\rm{c}}}{2}}\left(K-2c^{*}\right), & c^{*}\ge-K/2 \\
	 2\sqrt{Dg_{\rm{c}}K|c^{*}|}, & c^{*}<-K/2.
\end{array}
\right.
\label{eq:v_coop}
\end{equation}

\noindent Coordinates in the comoving reference frame~$(\tau, \zeta)$ are then defined in term of~$(t,x)$ as~$\tau=t$ and~$\zeta=x-v_{\rm{c}}t$. For $c^{*}\ge-K/2$, the shape of the wave profile is known exactly~\cite{murray:mathematical_biology,fife:allee_wave,aronson:allee_wave}:

\begin{equation}
	c(\zeta)=\frac{K}{1+e^{\sqrt{\frac{g_{c}}{2D}}K\zeta}}.
	\label{eq:profile_coop}
\end{equation}

\noindent For $c^{*}<-K/2$, the front has a qualitatively similar shape with~$c(\zeta)$ declining from~$1$ to~$0$, as~$\zeta$ goes from~$-\infty$ to~$+\infty$. The width of this transition scales as~$\sqrt{\frac{D}{g_{c}K|c^{*}|}}$, and~$c(\zeta)\sim e^{-\sqrt{\frac{g_{c}K|c^{*}|}{D}}\zeta}$ for large~$\zeta$~\cite{murray:mathematical_biology,fisher:wave,kolmogorov:wave}.

\subsection*{Evolutionary dynamics}

Similar to organisms, alleles encoding cooperator or defector phenotypes can spread in populations. Under the assumption of short-range and isotropic migration, the spreading of genetic changes can also be modeled by a reaction-diffusion equation: 

\begin{equation}
	\frac{\partial f}{\partial t} = D\frac{\partial^{2} f}{\partial x^{2}} + G_{f}(f)f,
	\label{eq:dfdt_sp}
\end{equation}

\noindent provided the alleles do not affect how organisms migrate and disperse. Here,~$f(t,x)\in(0,1)$ is the frequency~(fraction) of defectors, and~$G_{f}(f)$ is the relative growth rate of defectors describing the force of frequency-dependent natural selection. Note that the diffusion constant in equation~(\ref{eq:dfdt_sp}) is the same as in equation~(\ref{eq:dcdt_sp}) because both genetic and population spreading are due to the same migration process.  

The following model of frequency-dependent selection is most commonly used because of its simplicity and because it appears naturally as a weak-selection limit of evolutionary game theory~\cite{nowak:book,korolev:mutualism,frey:review}

\begin{equation}
	G_{f}(f)f = g_{f}f(1-f)(f^{*}-f),
	\label{eq:dfdt_wm}
\end{equation}

\noindent where~$g_{f}\ge0$ is the strength of selection, and~$f^{*}$ is the preferred or equilibrium frequency of defectors. In spatially homogeneous populations, cooperators and defectors coexist at a stable fixed point~$f=f^{*}$, provided~$f^{*}\in(0,1)$. The coexistence of cooperators and defectors has been observed experimentally; see, e.g., \cite{gore:snow_drift}. When~$f^{*}>1$, defectors outcompete cooperators, while cooperators prevail when~$f^{*}<0$. Negative~$g_{f}$ and~$f^{*}\in(0,1)$ describe a bistable behavior, e.g. due to a chemical warfare, and is not considered here. In game theory, these four scenarios are known as the snowdrift, prisoner's dilemma, harmony, and coordination games respectively~\cite{frey:review}.  

To understand the maintenance of cooperation, we need to know how defectors invade cooperators; see figure~\ref{fig:three_velocities}cd. The velocity of this invasion can be calculated using the results of~\cite{kolmogorov:wave} and is given by

\begin{equation}
	v_{\rm{i}}=2\sqrt{DG_{f}(0)}=2\sqrt{Dg_{f}f^{*}}.
	\label{eq:v_inv}
\end{equation}

\noindent In the comoving reference frame, the frequency of defectors changes from~$0$ to~$f^{*}$ (or from~$0$ to~$1$, if~$f^{*}>1$), as~$\zeta$ goes from~$-\infty$ to~$+\infty$. The characteristic width of this transition scales as~$\sqrt{D/(g_{f}f^{*})}$.

\subsection*{Coupling between ecological and evolutionary dynamics}

In the preceding discussion of population and genetic waves, we avoided the coupling between ecology and evolution either by assuming a constant genetic composition (no defectors) or neglecting the changes in population properties (e.g. carrying capacity) due to defector invasion. More general situations require a combined model with~$G_{c}$ and~$G_{f}$ depending on both~$c$ and~$f$. Here, we consider a natural extension of equations~(\ref{eq:dcdt_wm}) and~(\ref{eq:dfdt_wm}):

\begin{gather}
\begin{aligned}
	& G_{c}(c,f)c = g_{c}(f)c[K(f)-c][c-c^{*}(f)] \\
	& G_{f}(c,f)f = g_{f}(c)f(1-f)[f^{*}(c)-f],
\end{aligned}
	\label{eq:cf_wm}
\end{gather}

\noindent where the parameters of population dynamics depend on the genetic composition, and the parameters of evolutionary dynamics depend on population density.

Two comments on the functional form of our model are in order. First, equation~(\ref{eq:cf_wm}) allows for the most general dynamics that reduces to the classic models of frequency-dependent selection and cooperative growth with an Allee effect. Second, $G_{c}(c,f)$ and~$G_{f}(c,f)$ of an arbitrary functional form can be recast in the form of equation~(\ref{eq:cf_wm}) by allowing~$g_{c}$ to depend on~$c$ and~$g_{f}$ on~$f$. We expect these dependences to be small in many situations because the other terms in equation~(\ref{eq:cf_wm}) describe the most important aspects of the population dynamics. The analysis presented below, however, does not depend on the assumption that~$g_{c}$ is only a function of~$f$ and~$g_{f}$ in only a function of~$c$, provided~$g_{c}(c,f)>0$,~$g_{f}(c,f)>0$, and~$G_{f}$ is a decreasing function of~$f$ for~$f\in(0,f^{*})$. In the following, we will assume that~$G_{c}$ and~$G_{f}$ satisfy these conditions and will illustrate our results in the context of a simpler model given by equation~(\ref{eq:cf_wm}). The advantage of this approach is that many calculations can be carried out explicitly in the simpler model thus allowing us to provide an intuitive interpretation of the results.

We also note that equation (\ref{eq:cf_wm}) is phenomenological in nature and is not derived from a more mechanistic description of species interactions. One the one hand, this approach allows us to present a very general analysis that is valid for a large number of populations. On the other hand, our model cannot answer more mechanistic questions, e.g., how the dynamics of~$f^{*}(c)$ are constrained in any particular population, or how an increase in the death rate affects the evolutionary dynamics. 

The behavior of~$g_{c}(f)$ and~$g_{f}(c)$ has not been previously investigated; therefore, we typically assume that these two functions are constants. The dependencies of other model parameters in equation~(\ref{eq:cf_wm}) on~$c$ or~$f$ are known qualitatively from previous experimental and theoretical studies. Note, however, that most of our results are derived for general functional forms and do not depend on any specific assumptions about the parameters. 

Several experiments have confirmed the naive expectation that~$K(f)$ is a decreasing function~\cite{rainey:surface_film, turner:rna, griffin:pathogenic}. Interestingly, the opposite effect of defectors has also been observed in recent experiments with budding yeast, where it was established that mixed populations of defectors and cooperators have a higher carrying capacity than pure populations of cooperators~\cite{macLean:high_k}. For simplicity, we assume that~$K$ is a constant in most numerical solutions. 

The Allee threshold~$c^{*}$ is expected to increase with~$f$, and this was indeed observed both in models~\cite{hauert:prs_public_goods} and recent experiments~\cite{sanchez:yeast_dynamics}. In numerical solutions, we use~$c^{*}(f)=c^{*}(0)/(1-f)$, which is equivalent to requiring a certain density of cooperators to produce a sufficient amount of public goods for population growth.

As we show below, the dependence of the preferred frequency~$f^{*}$ on population density is particularly important for the cooperator-defector dynamics during range expansions. Modeling of public goods games revealed that, under certain conditions,~$f^{*}$ is a increasing function of~$c$~\cite{hauert:prs_public_goods, hauert:tpb_public_goods}, which has recently been confirmed by experiments with budding yeast~\cite{sanchez:yeast_dynamics}. For simplicity, in numerical solutions we assume that~$f^{*}$ changes from a high to a low value at a critical population density~$\bar{c}$; in other words, we use~$f^{*}(c)=f^{*}(0)+[f^{*}(1)-f^{*}(0)]\heaviside(c-\bar{c})$, where~$\heaviside(\cdot)$ is the Heaviside step function, which equals one for positive arguments and zero for negative arguments. Note that the value of~$f(0)$ can be negative when cooperators outcompete defectors at low population densities.

A spatial model with~$G_{c}$ and~$G_{f}$ defined in equation~(\ref{eq:cf_wm}) cannot be obtained by simply combining equations~(\ref{eq:dcdt_sp}) and~(\ref{eq:dfdt_sp}) because changes in population density and defector frequency due to migration are not independent. Instead, migration terms have to be added to the dynamics of cooperator and defector densities, defined as~$c_{\rm{c}}=c(1-f)$ and~$c_{\rm{d}}=cf$. Upon using equation~(\ref{eq:cf_wm}) to calculate the growth rates for~$c_{\rm{c}}$ and~$c_{\rm{d}}$ and adding the diffusion terms, we obtain that the spatial dynamics of mixed populations can be described by the following equation

\begin{gather}
\begin{aligned}
	& \frac{\partial c_{\rm{c}}}{\partial t} = D\frac{\partial^{2} c_{\rm{c}}}{\partial x^{2}} + G_{c}c_{\rm{c}} - G_{f}c_{\rm{d}} \\
	& \frac{\partial c_{\rm{d}}}{\partial t} = D\frac{\partial^{2} c_{\rm{d}}}{\partial x^{2}} + G_{c}c_{\rm{d}} + G_{f}c_{\rm{d}}. \\
\
\end{aligned}
	\label{eq:cd_sp}
\end{gather}

\noindent Here, the terms with~$G_{c}$ come from the ecological dynamics controlling the population density, and the terms with~$G_{f}$ come from the evolutionary dynamics controlling the relative abundances of cooperators and defectors. Note that we used the same diffusion constants for both cooperators and defectors. This is justified as long as the mutations causing defector phenotype do not affect dispersal. 

\begin{figure}[!ht]
\begin{center}
\includegraphics{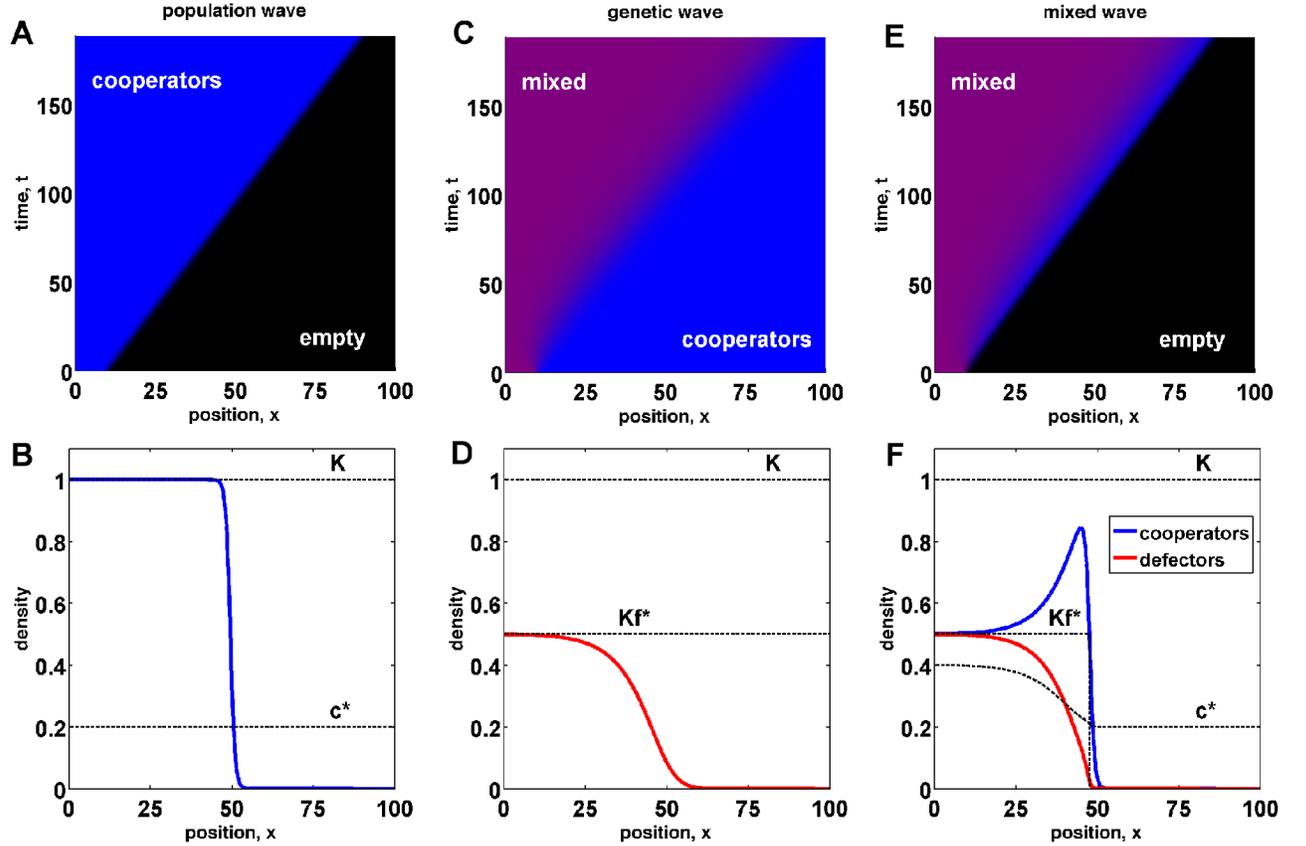}
\end{center}
\caption{{\bf Population, genetic, and mixed traveling waves.} (a) A population of cooperators~(blue) expands into an empty territory~(black) with a constant velocity~$v_{\rm{c}}$. (b) shows the density profile of the population wave in~(a). (c)  Defectors invade cooperators establishing a mixed population. The color encodes defector frequency with blue corresponding to pure cooperators and red to pure defectors. Mixed populations then have a magenta color. The invasion wave moves at a constant velocity~$v_{\rm{i}}$. (d) shows the defector density profile of a genetic wave in~(c). (e) A mixed population of defectors and cooperators expands into empty territory with a constant velocity~$v_{\rm{m}}$. Initially, cooperators and defectors are homogeneously mixed in equal proportion. As the population expands, the relative frequency of cooperators increases at the wave front, but stays constant in the interior of the population. (f) The density profiles of defectors and cooperators show the enhancement of cooperation at the expansion front and a small lag between defectors and cooperators. For all panels in this figure, we used~$K=1$, $c^{*}(f)=0.2/(1-f)$, $f^{*}(c)=5\heaviside(c-0.5)-4.5$, $g_{c}=2$, $g_{f}=0.25$, $D=0.5$, and numerically solved equation~(\protect{\ref{eq:cd_sp}}). The wave profiles are plotted for~$t=94$.}
\label{fig:three_velocities}
\end{figure}

A range expansion of a mixed population is shown in figure~\ref{fig:three_velocities}ef. Similar to the pure cooperator and invasion waves, mixed waves can spread with a constant velocity~$v_{\rm{m}}$ after an initial transient. Both cooperator and defector density profiles reach a constant shape in the comoving reference frame, but these shapes are different. In particular, the relative frequency of cooperators is much higher at the front than in the population bulk. This behavior is expected because~$f^{*}$ is a decreasing function of~$c$, and population density declines at the front. 

It is convenient to separate evolutionary and ecological dynamics, so we recast equation~(\ref{eq:cd_sp}) in terms of population density and defector frequency:

\begin{gather}
\begin{aligned}
	& \frac{\partial c}{\partial t} = D\frac{\partial^{2} c}{\partial x^{2}} + G_{c}(c,f)c \\
	& \frac{\partial f}{\partial t} = D\frac{\partial^{2} f}{\partial x^{2}} + 2D\frac{\partial \ln(c)}{\partial x}\frac{\partial f}{\partial x} + G_{f}(c,f)f \\
\end{aligned}
	\label{eq:cf_sp}
\end{gather}

\noindent Note that the additional advection term~$2D\frac{\partial \ln(c)}{\partial x}\frac{\partial f}{\partial x}$ appears because changes in~$f(t,x)$ due to migration depend on~$c(t,x)$. In particular, migration increases the relative frequency of organisms in low-density regions if these organisms are more abundant in nearby high-density regions because regions with higher density send out more migrants. This effect of density gradients is particularly important at the wave front, where~$c(t,x)$ is changing rapidly, and there is a constant imbalance between migrants coming from the high-density colonized regions and migrants coming from low-density uncolonized regions. A more general discussion of the advection terms in populations with many different species (or alleles) can be found in~\cite{vlad:hydrodynamics}. 

\subsection*{Numerical solutions and simulations}

Numerical solutions of equation~(\ref{eq:cd_sp}) can be easily obtained by standard methods~\cite{press:numerical_recipes}. Equations~(\ref{eq:dcdt_sp}) and~(\ref{eq:dfdt_sp}) did not require a separate solver because they are special cases of equation~(\ref{eq:cd_sp}). We used an explicit forward-time centered-space (FTCS) finite-difference method (4-point stencil)~\cite{press:numerical_recipes}. Spatial discretization step was~$0.1$, and temporal discretization step was~$\min\{10^{-3}/D,10^{-2}/\max\{g_{f}\max\{|f^{*}|,1\},g_{c}K^{2}\}\}$. This level of discretization was sufficient to ensure that numerical wave velocities did not differ from the analytical results in equations~(\ref{eq:v_coop}) and~(\ref{eq:v_inv}) by more than a percent. For cooperator and mixed waves, the initial population typically occupied the left~$10$\% of the habitat. For genetic waves, the whole habitat was occupied by cooperators, while defectors were initially present only the left~$10$\% of the habitat. At~$t=0$, populations were always in local equilibrium, i.e. with~$c=K(f)$ and~$f=f^{*}(c)$. We used no-flux boundary conditions and computed solutions up to the time when the wave had spread into~$90$\% of the habitat. The habitat size~$L$, i.e. the distance between the left and right boundaries, was typically equal to~$100$.

We also performed individual-based simulations to demonstrate the possibility of stochastic splitting. The simulations were done on a one-dimensional lattice of sites each with a carrying capacity~$K$. Every time step consisted of one possible migration event and one reproduction or death event at every site. During a migration event, a randomly chosen organism migrated with probability~$m$ to one of the two neighboring sites. Migration was isotropic, and we imposed no-flux boundary conditions. During a reproduction or death event, the population at a site could increase by one, decrease by one, or remain unchanged. The probability of birth was given by~$g_{c}c^{2}(K+c^{*})/K^{3}/(1+g_{c}c^{2}(K+c^{*})/K^{3}+g_{c}(c^{3}+cKc^{*})/K^{3})$, and the probability of death was given by~$g_{c}(c^{3}+cKc^{*})/K^{3}/(1+g_{c}c^{2}(K+c^{*})/K^{3}+g_{c}(c^{3}+cKc^{*})/K^{3})$. Death events were equally likely to eliminate cooperators and defectors, but birth events created either a new cooperator or a new defector with probabilities determined by the interaction matrix~$A$ which was chosen to mimic equation~(\ref{eq:cf_wm}). Given that a birth event occurred, a cooperator was born with probability~$(1-f)(A_{\rm{cc}}(1-f)+A_{\rm{cd}}f)/(A_{\rm{cc}}(1-f)^{2}+A_{\rm{cd}}(1-f)f+A_{\rm{dc}}f(1-f)+A_{\rm{dd}}f^{2})$, where~$f$ is the frequency of defectors, and we used letters~$\rm{c}$ and~$\rm{d}$ as indices. Otherwise, a defector was born. $K$~time steps constituted a generation.

\section*{Results}

In the previous section, we formulated a reaction-diffusion model that includes the coupling between ecological and evolutionary dynamics of cooperators and defectors. This model, equation~(\ref{eq:cf_sp}), and its special cases, equations~(\ref{eq:dcdt_sp}) and~(\ref{eq:dfdt_sp}), describe range expansions of cooperators with velocity~$v_{\rm{c}}$, invasion of cooperators by defectors with velocity~$v_{\rm{i}}$, and spreading of mixed waves of defectors and cooperators with velocity~$v_{\rm{m}}$. For mixed waves, we find that the frequency of cooperators is higher at the wave front because cooperators are favored at low population densities; see figure~\ref{fig:three_velocities}ef. In this section, we show that, under certain conditions, cooperators can take over the entire wave front and split from defectors by colonizing new territories faster than defectors can invade from behind. We also investigate how the speed of mixed waves depends on the parameters of the model and show that, when evolutionary dynamics are much slower than ecological dynamics~($g_{f} \ll g_{c}K^{2}$), mixed waves can experience long periods of acceleration. 

\begin{figure}
\begin{center}
\includegraphics{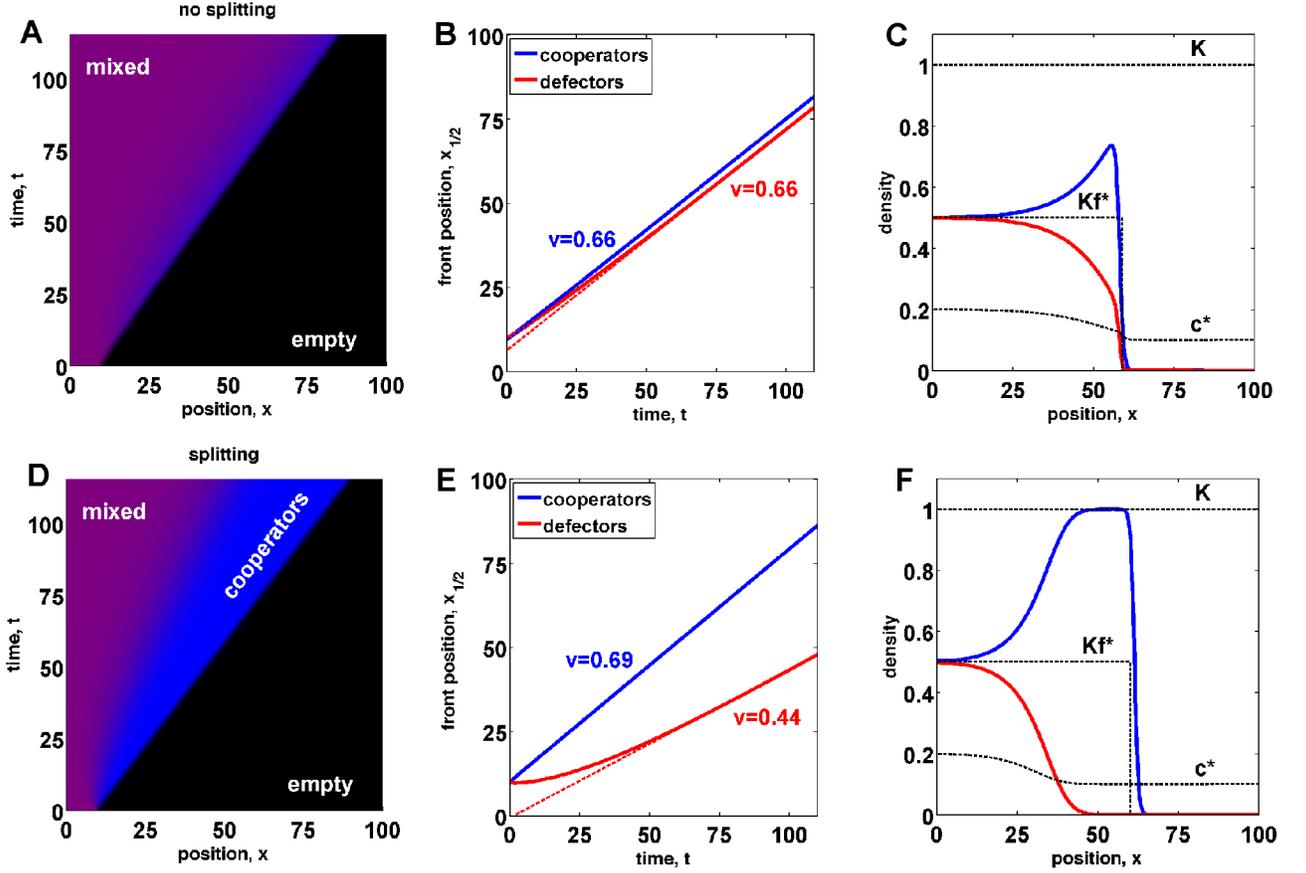}
\end{center}
\caption{{\bf Splitting of cooperators from defectors}. The top panels (a), (b), and (c) show a mixed wave where cooperators and defectors spread together. The bottom panels (d), (e), and (f) show cooperators splitting from defectors and colonizing new territories faster than defectors can invade them from behind. For all panels in this figure, we used~$K=1$, $c^{*}=0.1/(1-f)$, $g_{c}=3$, $g_{f}=0.25$, $D=0.5$, and numerically solved equation~(\protect{\ref{eq:cd_sp}}). The difference between the top and the bottom panels is in~$f^{*}(c)$. No splitting occurs for~$f^{*}=5\heaviside(c-0.2)-4.5$, in the top panels, but a higher value of~$\bar{c}$ ensures splitting in the bottom panels, where~$f^{*}=5\heaviside(c-0.9)-4.5$. (a) and (d) show a mixed wave expanding into new territories. In both panels, the front has a higher frequency of cooperators~(blue) compared to the population bulk~(magenta). However, the size of the region enriched in cooperators remains constant in the top panel, while it grows linearly with time in the bottom panel. (b) and (e) show the positions of the expansion fronts for cooperators and defectors as a function of time. The position of the front is defined as the rightmost point where the density of cooperators or defectors reaches half of its value in the bulk. The asymptotic rates of expansions are shown with dashed lines. From~(e), we know the velocity of cooperators~$v_{\rm{c}}=0.69$ and the invasion velocity~$v_{\rm{i}}=0.44$. These velocities are the same as in~(b) because they do not depend on~$\bar{c}$. Note that the mixed velocity~$v_{\rm{m}}=0.66$ is smaller than~$v_{\rm{c}}$ because defectors increase the Allee threshold and slow down the population. More importantly, defectors, which are also spreading with velocity~$v_{\rm{m}}$, colonize new territories faster than they can invade cooperators~($v_{\rm{m}}>v_{\rm{i}}$). (c) and (f) show density profiles of defectors and cooperators at~$t=75$. As expected, the lag between cooperators and defectors is larger in (f), where there is a region of pure cooperators at the carrying capacity.}
\label{fig:splitting}
\end{figure}

The behavior of mixed waves depends on the ratio of cooperator and invasion velocities. When~$v_{\rm{i}}>v_{\rm{c}}$, defectors invade faster than cooperators can spread into new territories; therefore, any initial condition leads to a mixed population of defectors and cooperators. This mixed population will expand into empty territories with both defectors and cooperators spreading at the same velocity~$v_{m}$. When~$v_{\rm{c}}>v_{\rm{i}}$, two qualitatively different outcomes are possible. Either cooperators and defectors can spread together in a mixed wave with velocity~$v_{\rm{m}}$~(figure~\ref{fig:splitting}abc), or a mixed wave can split into an ecological expansion wave and an evolutionary invasion wave~(figure~\ref{fig:splitting}def). In the latter scenario, the population expansion with velocity~$v_{\rm{c}}$ is driven solely by cooperators followed by a slower invasion by defectors with velocity~$v_{\rm{i}}$. 

The analysis of equation~(\ref{eq:cf_sp}) is complicated both by the coupling between the two differential equations and by the unknown dependences of the parameters on population density and defector frequency. To reduce this complexity, we first consider the effect of evolution on ecology and the effect of ecology on evolution separately and then analyze the general case.

\subsection*{Effects of evolutionary on ecological dynamics}

By neglecting the effect of ecology on evolution and setting~$g_{f}(c)=g_{f}$ and~$f^{*}(c)=f^{*}$, we can immediately see that~$f(t,x)=f^{*}$ is a stationary solution of equation~(\ref{eq:cf_sp}). With this solution for~$f(t,x)$, the dynamics of~$c(t,x)$ become identical to that in equation~(\ref{eq:dcdt_sp}) with the parameters~$c^{*}$,~$K$,~and~$g_{c}$ evaluated at~$f=f^{*}$. Therefore, the velocity and density profile of the mixed wave can be immediately obtained from equations~(\ref{eq:v_coop}) and~(\ref{eq:profile_coop}), e.g., the mixed velocity~$v_{\rm{m}}$ is given by

\begin{equation}
	v_{f^{*}}=
\left\{
\begin{array}{ll}
	 \sqrt{\frac{Dg_{\rm{c}}(f^{*})}{2}}\left[K(f^{*})-2c^{*}(f^{*})\right], & c^{*}(f^{*})\ge-K/2 \\
	 2\sqrt{Dg_{\rm{c}}(f^{*})K(f^{*})[-c^{*}(f^{*})]}, & c^{*}(f^{*})<-K/2.
\end{array}
\right.
\label{eq:v_fs}
\end{equation}

\noindent Defectors are expected to decrease the rate of spatial expansions~\cite{hilker:pathogens}. Consistent with this expectation, equation~(\ref{eq:v_fs}) predicts that~$v_{f^{*}}<v_{\rm{c}}$, provided that all or some of the growth parameters~($g_{c}$,~$K$, and~$c^{*}$) substantially decrease with~$f$. Note that, because~$f^{*}$ is an attracting fixed point everywhere in space, equation~(\ref{eq:cf_sp}) does not allow wave splitting when evolutionary dynamics are decoupled from population dynamics.

\subsection*{Effects of ecological on evolutionary dynamics}

To understand the effect of ecology on evolution, we assume that the ecological parameters do not depend on the frequency of defectors:~$K(f)=K$,~$g_{c}(f)=g_{c}$, and~$c^{*}(f)=c^{*}$. The partial differential equation for~$c(t,x)$ is then independent from the dynamics of~$f(t,x)$, so the velocity of the population wave and wave profile are given by equations~(\ref{eq:v_coop}) and~(\ref{eq:profile_coop}) respectively. As a result, equation~(\ref{eq:cf_sp}) is reduced to a single equation for the defector frequency~$f(t,x)$ by substituting the solution for~$c(t,x)$. In the reference frame~$(\tau,\zeta)$ comoving with the population wave, this substitution results in an explicit dependence of the growth dynamics on~$\zeta$ through the~$c$-dependence of~$G_{f}$ and~$2D\frac{\partial}{\partial x}\ln(c)\frac{\partial f}{\partial x}$. Compared to equation~(\ref{eq:dfdt_sp}), the additional term~$2D\frac{\partial}{\partial x}\ln(c)\frac{\partial f}{\partial x}$ comes from the effect of density gradients on spatial diffusion of alleles. This term is similar to advection in reaction-diffusion equations with the medium moving at a velocity~$v_{\rm{a}}=-2D\frac{\partial}{\partial x}\ln(c)$. To make this analogy explicit, we need to take the advection velocity~$v_{\rm{a}}$ inside the partial derivative, which results in the following equation for~$f(\tau,\zeta)$

\begin{equation}
	 \frac{\partial f}{\partial \tau} = D\frac{\partial^{2} f}{\partial \zeta^{2}} + \frac{\partial}{\partial \zeta}[(v_{\rm{c}}-v_{a})f]+ [G_{f} + \frac{\partial v_{\rm{a}}}{\partial \zeta}]f,
	\label{eq:f_cm}
\end{equation}

\noindent where the term with~$v_{\rm{c}}$ appears because we changed to the comoving reference frame. From equation~(\ref{eq:f_cm}), one can see that~$2D\frac{\partial}{\partial x}\ln(c)\frac{\partial f}{\partial x}$ results in both effective advection and effective growth. The effective terms are only functions of~$\zeta$ and can be computed from the known density profile~$c(\zeta)$. For~$c^{*}>-K/2$, this computation can be carried out explicitly with the following results  

\begin{gather}
\begin{aligned}
	& v_{\rm{a}} \equiv -2D\frac{\partial}{\partial x}\ln(c) = \sqrt{2Dg_{\rm_{c}}}K(1-c/K) \\
	& g_{\rm{e}} \equiv \frac{\partial v_{\rm{a}}}{\partial \zeta} = g_{c}c(K-c).
\end{aligned}
	\label{eq:effective}
\end{gather}

\noindent The effective growth term is always positive and peaks at the middle of the population wave front where~$c=K/2$ and~$\zeta=0$~(by definition). The effective advection term is also positive, provided~$f$ decreases with~$\zeta$, which is expected when cooperators are favored at low population densities. As we show below and in Text S1, these two terms in some cases allow defectors to keep pace with cooperators even when~$v_{\rm{i}}<v_{\rm{c}}$.

The existence of a mixed traveling wave, where both cooperators and defectors spread at the same velocity, is equivalent to the existence of a steady state for~$f(\tau,\zeta)$ in equation~(\ref{eq:f_cm}). When a steady state does not exist, population and genetic waves split, and~$f(\tau,\zeta)$ shifts to negative~$\zeta$ with velocity~$v_{\rm{c}}-v_{\rm{i}}$. Since~$G_{f}$ decreases with~$f$, the existence of a steady state requires that operator~$\mathcal{L}$, obtained by linearizing the right hand side of equation~(\ref{eq:f_cm}) with respect to~$f$, has a positive eigenvalue~$\lambda$. Indeed, if all eigenvalues of~$\mathcal{L}$ are negative, then~$\frac{\partial f}{\partial \tau}<0$ because~$G_{f}(c,f)<G_{f}(c,0)$, while a positive eigenvalue ensures that small~$f(\tau,\zeta)$ will increase until~$G_{f}(c,f)$ is sufficiently diminished so that~$\frac{\partial f}{\partial \tau}=0$. Thus, we look for a solution of the following equation with~$\lambda>0$

\begin{equation}
	Df''(\zeta)+[v_{\rm{c}}-v_{\rm{a}}(\zeta)]f'(\zeta)+g_{\rm{l}}(\zeta)f(\zeta) = \lambda f(\zeta),
	\label{eq:eigen}
\end{equation}

\noindent where~$g_{\rm{l}}(\zeta)=G_{f}(c(\zeta),0)$, and we used primes to denote derivatives with respect to~$\zeta$. Equation~(\ref{eq:eigen}) can be transformed to a canonical form by the following change of variables

\begin{gather}
\begin{aligned}
	& f(\zeta) = e^{-u(\zeta)}\psi(\zeta), \;\;\mbox{where}\\
	& u(\zeta) = \int\frac{v_{\rm{c}}-v_{\rm{a}}(\zeta)}{2D}d\zeta, \\
\end{aligned}
	\label{eq:eigen_change}
\end{gather}

\noindent which also insures that~$\psi(\pm\infty)=0$; see Text S1. The result reads

\begin{gather}
\begin{aligned}
	& -D\psi''+V(\zeta)\psi = -\lambda\psi, \;\;\mbox{where}\\
	& V(\zeta) = -g_{\rm{l}}(\zeta)-\frac{1}{2}v_{\rm{a}}'(\zeta)+\frac{1}{4D}[v_{\rm{c}}-v_{\rm{a}}(\zeta)]^{2}. \\
\end{aligned}
	\label{eq:eigen_canonical}
\end{gather}

\noindent We can now use the standard theory of second order differential equations to establish conditions necessary for the existence of a solution for~$\lambda>0$; see~\cite{titchmarsh:second_order_ode, merzbacher:qm}. These conditions require that there must be a sufficiently large region where the potential~$V(\zeta)$ is negative and the values of~$V(\zeta)$ in this region must be sufficiently low. To show this, we multiply both sides of equation~(\ref{eq:eigen_canonical}) by~$\psi$ and integrate over~$\zeta$, which gives

\begin{equation}
D\int_{-\infty}^{\infty}\psi'^{2}(\zeta)d\zeta + \lambda\int_{-\infty}^{\infty}\psi^{2}(\zeta)d\zeta + \int_{-\infty}^{\infty}V(\zeta)\psi^{2}(\zeta)d\zeta = 0
\label{eq:variation_principle}
\end{equation}

\noindent after an integration by parts in the first term. Since the first two terms in equation~(\ref{eq:variation_principle}) are positive, the third term must be negative, which in turn requires that there exists a region where~$V(\zeta)<0$ because~$\psi^{2}(\zeta)\ge0$. This region has a finite width because~$V(\pm\infty)>0$, which follows from equations~(\ref{eq:cf_wm}) and~(\ref{eq:effective}). Indeed, for~$\zeta\to-\infty$, we find that~$V(-\infty)=-g_{f}(K)f^{*}(K)+v_{\rm{c}}^{2}/(4D)=(v_{\rm{c}}^{2}-v_{\rm{i}}^{2})/(4D)$, which is positive since~$v_{\rm{c}}>v_{\rm{i}}$. In agreement with the earlier discussion, when~$v_{\rm{c}}<v_{\rm{i}}$, the potential at~$-\infty$ is negative ensuring the existence of eigenfunction for~$\lambda>0$ and, therefore, the existence of mixed waves. For~$\zeta\to+\infty$, we find that~$V(+\infty)=-g_{f}(0)f^{*}(0)+[v_{\rm{c}}-v_{\rm{a}}(+\infty)]^{2}/(4D)$, which is positive as long as defectors are sufficiently disfavored at low densities, e.g. when~$f^{*}(0)<0$. Thus,~$V(\zeta)$ is a potential well, and the existence of an eigenfunction for~$\lambda>0$ requires that this potential well be sufficiently wide and deep.

We now interpret the effects of the three terms on the right hand side of equation~(\ref{eq:eigen_canonical}) in the context of the depth and width of the potential well~$V(\zeta)$. The first term lowers~$V(\zeta)$ for~$c>\bar{c}$ in the population bulk~($\zeta\to-\infty$), but it increases~$V(\zeta)$ for~$c<\bar{c}$ at the front~($\zeta\to+\infty$). The second term is always negative; it deepens the potential well around~$\zeta=0$ and vanishes for~$\zeta\to\pm\infty$. The third term is always positive, but the reduction in the depth of the potential well due to~$v_{\rm{c}}$ is lessened by~$v_{\rm{a}}$~(at least for some~$\zeta$). The transition from non-splitting to splitting behavior can then be achieved by reducing the potential well. In particular, the transition threshold can be crossed by increasing~$\bar{c}$, the density at which natural selection starts to favor cooperators over defectors, as shown in figures~\ref{fig:splitting} and~\ref{fig:generality}a. Decreasing~$f^{*}(0)$ or~$v_{\rm{i}}/v_{\rm{c}}$ has a similar effect. 


To understand wave splitting better, we solve a special case of equation~(\ref{eq:eigen}) exactly in Text S1 and find how the threshold between splitting and non-splitting behavior depends on model parameters. An exact solution can be found when~$c^{*}=0$, and~$f^{*}(c)=f^{*}$ for~$c>\bar{c}$, but defectors are not viable for~$c<\bar{c}$. In other words, we impose an absorbing boundary condition at~$\bar{\zeta}$ such that~$c(\bar{\zeta})=\bar{c}$. With these simplifications, we find that splitting occurs provided 

\begin{equation}
	\frac{\bar{c}}{K} > \frac{1-\sqrt{1-v_{\rm{i}}^{2}/v_{\rm{c}}^{2}}}{2}.
	\label{eq:splitting_condition}
\end{equation}

We draw three important conclusions from this result. First, the exact solution confirms that both splitting and non-splitting behaviors are possible depending on the parameters of the model. Second, the severity of selection against defectors required for splitting increases as~$v_{\rm{i}}$ approaches~$v_{\rm{c}}$. Third, defectors can spread faster in mixed waves than they can invade cooperators when splitting does not occur. These three conclusions do not depend on the simplifying assumptions used to derive equation~(\ref{eq:splitting_condition}); see the discussion below and figures~\ref{fig:splitting} and~\ref{fig:generality}a. We now discuss the biological significance of wave splitting. 

The possibility of wave splitting has important implications for the evolution of cooperation. When splitting is possible, cooperators outrun defectors, leading to a constant increase in the relative abundance of cooperators for as long as there are uncolonized territories. Frequent local extinctions followed by recolonizations could therefore maintain high levels of cooperation in natural populations. Equation~(\ref{eq:splitting_condition}) also suggests that traits that have little effect on the dynamics in well-mixed populations might be under selection during range expansions. One such trait is~$g_{f}$. In well-mixed populations, it only determines the rate of approach to the equilibrium not the equilibrium itself, while, in expanding populations,~$g_{f}$ affects the invasion velocity and the conditions for wave splitting, both of which are expected to be under selection. The critical density~$\bar{c}$ is another examples. It has little effect in the well-mixed populations, but large~$\bar{c}$ allows cooperators to escape from defectors during range expansions.      

The possibility of non-splitting highlights the importance of the coupling between ecology and evolution in the maintenance of genetic or species diversity during range expansions. Quite surprisingly, we find that defectors can spread in a mixed wave faster than they can invade cooperators despite the fact that defectors are disfavored or even eliminated by natural selection at the front. In other words, the community of cooperators and defectors is able to move to new territories while preserving the coexistence between the two phenotypes. Our results could, therefore, have important implications for conservation efforts to preserve genetic diversity or ecosystem integrity during potentially rapid range shifts due to climate change or habitat deterioration. For negative frequency-dependence discussed here, we found that diversity is more stable than one would naively predict from just measuring~$v_{\rm{c}}$ and~$v_{\rm{i}}$. Other types of interactions could be less resilient and would require managed interventions to prevent splitting. In Text S1, we show that interventions increasing advection or relative growth rate at the front can achieve that goal. 

\subsection*{General case of eco-evolutionary feedback}

\begin{figure}
\begin{center}
\includegraphics{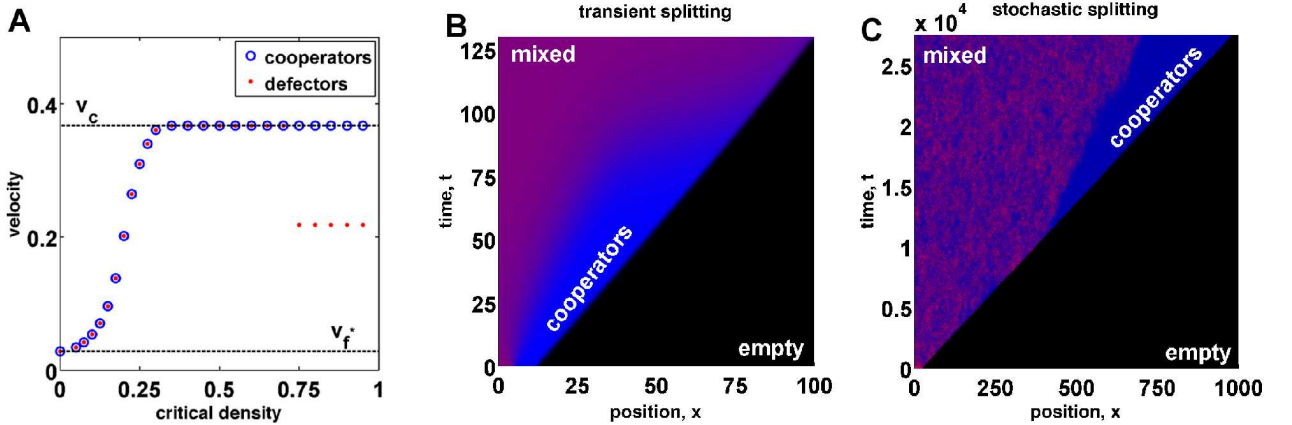}
\end{center}
\caption{{\bf Wave splitting under different conditions}. (a) shows how the velocities of cooperators~(blue circles), defectors~(red dots) depend on the critical density,~$\bar{c}$. For small~$\bar{c}$, the population moves with the velocity~$v_{f^{*}}$. As~$\bar{c}$ increases, the expansion first reaches velocity~$v_{\rm{c}}$ and then splits into a population and a genetic wave, which occurs when the velocities of cooperators and defectors are no longer the same. At this point, the velocity of cooperators and the population expansion is~$v_{\rm{c}}$, while the velocity of defectors is~$v_{\rm{i}}$. For this panel, we used~$K=1$,~$g_{c}=2$, $g_{f}=0.05$, $c^{*}(f)=0.24/(1-f)$, $f^{*}=10\heaviside(c-\bar{c})-9.5$, and~$D=0.5$. We used large habitat lengths up to~$1800$ because, close to the splitting transition, transient dynamics decay slowly. (b) Creation of a region of pure cooperators in front of a mixed population leads to transient splitting. Initially, cooperators seem to outrun defectors because~$v_{\rm{c}}>v_{\rm{i}}$. At later times, however, defectors catch up because the conditions for deterministic splitting are not met, and a small number of defectors that migrated to the front can follow cooperators and grow until a stable mixed wave is formed. For this panel, we used~$K=1$,~$g_{c}=3$, $g_{f}=0.25$, $c^{*}(f)=0.1/(1-f)$, $f^{*}=5\heaviside(c-0.1)-4.5$, and~$D=0.5$. (c) Stochastic splitting is possible even when the conditions for deterministic splitting are not satisfied, but~$v_{\rm{c}}>v_{\rm{i}}$. The waiting time for stochastic splitting can however be very long~(here~$10^{4}$ generations), even for moderately low population densities of~$100$ individuals per site used in this simulation, because splitting is caused by a rare fluctuation that creates a sufficiently large region of pure cooperators at the front. For this panel we used~$L=1000$, $m=0.1$, $K=100$, $g_{c}=0.05$, $c^{*}=0$, and the entries of the interaction matrix are given by~$A_{\rm{cc}}=2.5$, $A_{\rm{cd}}=1.5$, $A_{\rm{dc}}=2.65$ and $A_{\rm{dd}}=1$.}
\label{fig:generality}
\end{figure}

In the general case when ecology and evolution affect each other, the nature of the transition from non-splitting to splitting behavior remains the same. In particular, the condition for splitting is still given by the existence of a solution of equation~(\ref{eq:eigen}) with~$\lambda>0$ because, close to the splitting transition, the expansion front is populated almost exclusively by cooperators. To show this, let us consider how the dynamics at the front changes as one of the model parameters, say~$\bar{c}$, is varied so that the system behavior changes from non-splitting to splitting. Increasing~$\bar{c}$ directly increases the abundance of cooperators at the front and increases the velocity of the mixed wave from~$v_{f^{*}}$, when~$\bar{c}=0$, to~$v_{c}$, when splitting occurs. In addition to that, defectors lag more and more behind the front, as the splitting transition is approached. Indeed, close to the transition,~$G_{f}$ is barely sufficient for defectors to keep up with cooperators, and, since~$G_{f}$ decreases with~$f$, the fraction of defectors at the front must approach zero right before splitting occurs. These dynamics are illustrated in figure~\ref{fig:generality}a and further discussed in Text S1. Another effect of the coupling between ecology and evolution is that~$v_{\rm{i}}$ is no longer given by equation~(\ref{eq:v_inv}) and one has to solve equation~(\ref{eq:cf_sp}) to describe defector invasion. 

We would now like to discuss the effect of initial conditions on wave splitting. One may naively think that deterministic wave splitting can be achieved simply by creating a region of pure cooperators in front of a mixed population, even if populations do not split when started from a well-mixed state. Initially, the range expansion started from such an initial condition indeed resembles splitting with cooperators transiently outrunning defectors and increasing in relative abundance; see figure~\ref{fig:generality}b. Nevertheless, defectors eventually catch up because any initial condition~(with~$c(0,x)>0$ and~$f(0,x)>0$) has a nonzero projection on the eigenfunction of the operator~$\mathcal{L}$ with the largest positive eigenvalue. The growth of this projection ensures the establishment of defector population at the front even when~$v_{\rm{i}}<v_{\rm{c}}$ as shown in figure~\ref{fig:generality}b. The ability of defectors to catch up, however, relies on the assumption that~$c(t,x)$ and~$f(t,x)$ vary continuously and can increase from arbitrarily small values. When the discrete nature of organisms is taken into account, one finds that wave fronts have a finite width~\cite{hallatschek:fisher_wave}. Therefore, initial conditions can force splitting provided~$v_{\rm{c}}>v_{\rm{i}}$ and the region of pure cooperators is sufficiently large compared to the width of expansion and invasion wave profiles.

\begin{figure}
\begin{center}
\includegraphics{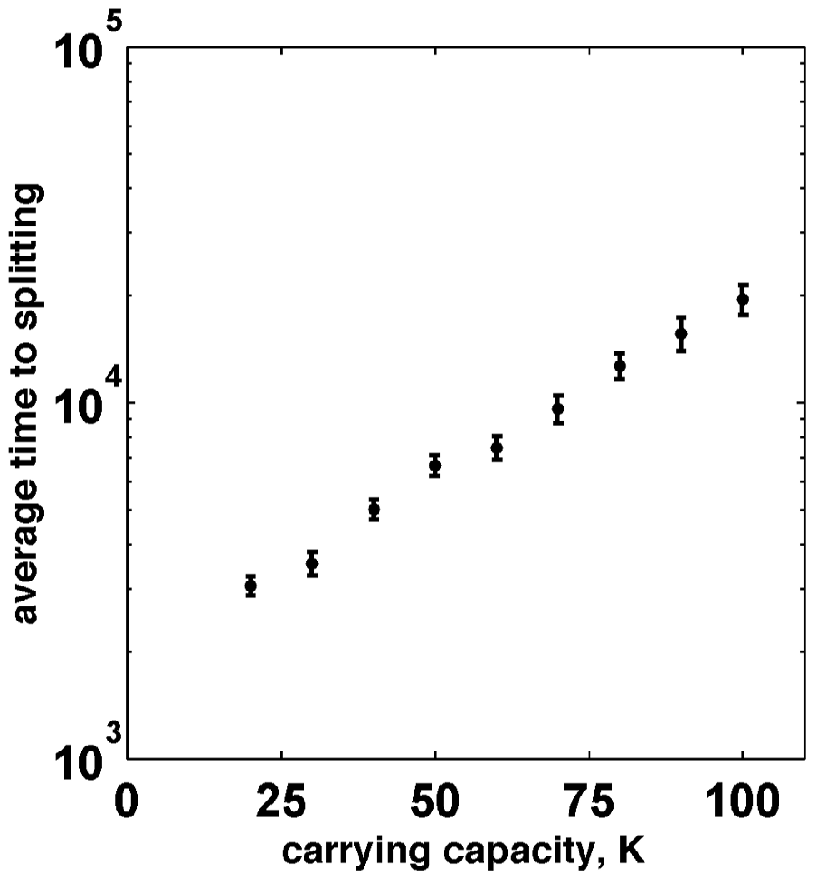}
\end{center}
\caption{{\bf The average time to stochastic splitting shows an exponential-like increase with the carrying capacity of the sites.} The parameters used in these simulations are the same as in figure~\protect{\ref{fig:generality}c} with the exception of~$L=2000$ for~$K<100$ and~$L=3000$ for~$K=100$. The error bars show the standard deviation of the mean. Similar increase in the average time to splitting was observed for increasing migration because larger~$m$ leads to longer wave profiles containing more organisms, and, therefore, smaller number fluctuations.}
\label{fig:time}
\end{figure}

Number fluctuations must also be considered for traveling waves of discrete entities. For biological systems, these fluctuations could arise due to random fluctuations of the environment or due to the randomness of births and deaths, which is known as genetic or ecological drift. Number fluctuations are largely irrelevant when~$v_{\rm{i}}>v_{\rm{c}}$ because defectors can always catch up with cooperators. In contrast, when~$v_{\rm{i}}<v_{\rm{c}}$, splitting is the ultimate outcome because, given enough time, a fluctuation will create a region of pure cooperators, which is sufficiently large to prevent defectors from catching up. This region of pure cooperators will then grow with time making it exceedingly unlikely for another fluctuation to destroy it. Although splitting is inevitable, it may take a very long time~(e.g.~$10^{4}$ generations in figure~\ref{fig:generality}c) because the deterministic dynamics discussed above create an effective activation barrier for stochastic splitting; see figure~\ref{fig:generality}c and~\ref{fig:time}. The magnitude of this barrier goes to zero as the conditions for deterministic splitting are approached. Stochastic splitting is also possible for frequency-independent selection, provided expansion and invasion velocities are different; see for example~\cite{hallatschek:life_front}.

In the preceding discussion, we neglected possible transitions between cooperators and defectors due to mutations or other heritable changes. This is justified on short time scales because mutation rates are typically small and even beneficial mutations struggle to survive number fluctuations~(genetic drift)~\cite{gillespie:textbook, korolev:review}. On long time scales, mutations will change the dynamics in two ways. First, the coexistence fraction~$f^{*}$ will be determined by both natural selection and the mutational pressures similar to the classic theory of two genetic variants~\cite{gillespie:textbook, korolev:review}. This shift in~$f^{*}$ does not change the dynamics qualitatively and could be included in our theory by modifying~$G_{f}(c,f)$ accordingly. The second and more important difference is that the region of pure cooperators will not expand indefinitely following a splitting event because defectors will appear in the interior of the region of pure cooperation due to a mutation in addition to invading the region of pure cooperators from behind. In the limit of rare mutations, the average length of the region of pure cooperators~$L_{\rm{pc}}$ will be determined by the balance between the time necessary to create this region after splitting and the time to the next successful defector mutation in the region of pure cooperators. The former time scales as~$L_{\rm{pc}}/(v_{\rm{c}}-v_{\rm{i}})$ for deterministic splitting. While the latter time scales as the inverse of the product of the mutation rate from cooperators to defectors~$\mu_{\rm{d}}$, population size~$K(0)L_{\rm{pc}}$, and fixation probability~$g_{f}[K(0)]f^{*}[K(0)]$, i.e. as~$\{\mu_{\rm{d}}L_{\rm{pc}}K(0)g_{f}[K(0)]f^{*}[K(0)]\}^{-1}$; see~\cite{gillespie:textbook, korolev:review}. For deterministic splitting, this balance yields~$L_{\rm{pc}}\approx\sqrt{(v_{\rm{c}}-v_{\rm{i}})/\{\mu_{\rm{d}}K(0)g_{f}[K(0)]f^{*}[K(0)]\}}$, while, for stochastic splitting,~$L_{\rm{pc}}$ will also be affected by the average waiting time to stochastic splitting. In the other limit of very frequent mutations, splitting would not be able to occur before additional defectors arise at the front, and the dynamics would be primarily determined by the balance between mutational transitions leading to mixed populations.

\begin{figure}
\begin{center}
\includegraphics{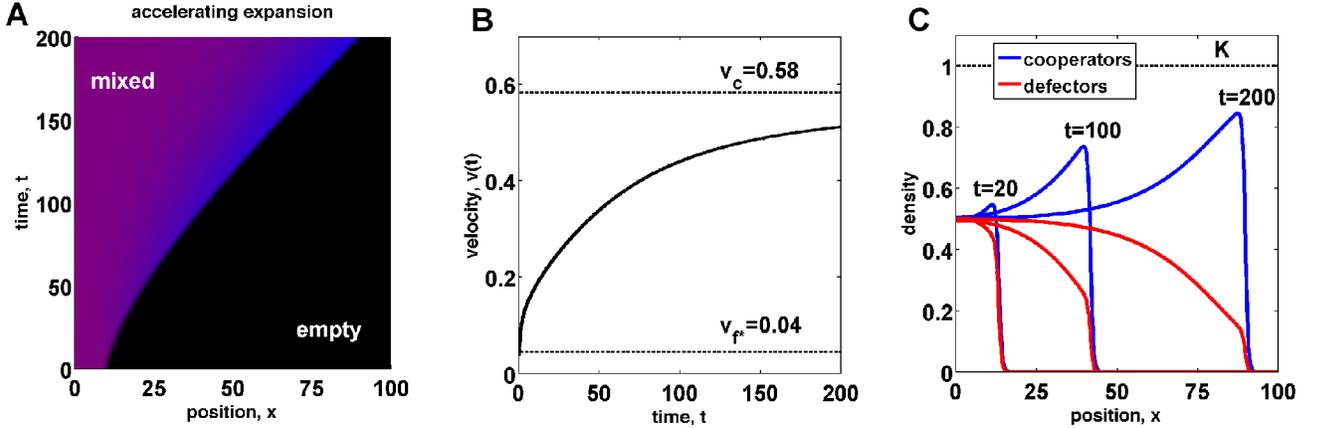}
\end{center}
\caption{{\bf Population waves can accelerate due to evolutionary changes at the front}. (a) A mixed wave of cooperators and defectors spreads into empty territories. The shape of this invasion is not linear in the~$(t,x)$ coordinates indicating wave acceleration. The acceleration is caused by a slow increase in the number of cooperators~(blue) at the front compared to the bulk~(magenta). (b) The velocity of the front increases from~$v_{f^{*}}$ to its asymptotic value~$v_{\rm{m}}$, which is smaller than~$v_{\rm{c}}$ because population and genetic waves do not split in this case. (c) Cooperator and defector density profiles at different times show both acceleration of the expansion and the increase in the frequency of cooperators at the front. For this figure, we used~$K=1$,~$g_{c}=5$, $g_{f}=0.1$, $c^{*}(f)=0.24/(1-f)$, $f^{*}=3\heaviside(c-0.5)-2.5$, and~$D=0.5$.}
\label{fig:acceleration}
\end{figure}

Another interesting consequence of the coupling between ecological and evolutionary dynamics is wave acceleration. We found that cooperators are favored at the wave front~(see figure~\ref{fig:three_velocities}c). As the frequency of cooperators is increasing at the front, the instantaneous velocity of the expansion wave must also increase because defectors slow down expansions~(see figure~\ref{fig:splitting}). The evolutionary change could however be very slow compared to the colonization dynamics leading to a gradual acceleration of the range expansion. Indeed, long periods of dramatic wave acceleration are possible in our model and are shown in figure~\ref{fig:acceleration}. The acceleration of expansion fronts has been observed in a number of species and is typically explained by the evolution of shorter generation times, greater dispersal abilities, or specific adaptations to the environment in the newly colonized regions~\cite{thomas:acceleration, phillips:toad_acceleration, phillips:toad_tropical,gray:worm_expansion}. Our analysis of cooperator and defector waves suggests that a changing rate of expansion could simply be a consequence of the slow adjustment of the genetic composition at the wave front to a new low-density optimum, which is different from that in the population bulk.

\clearpage

\section*{Discussion}

Understanding the link between ecology and evolution remains a major open problem~\cite{kokko:ecoevo}. This coupling is particularly important during range expansions, which are often accompanied by both demographic and evolutionary changes~\cite{phillips:toad_acceleration, phillips:toad_tropical, gray:worm_expansion, shine:spatial_sorting}. We formulated a simple two-allele model that is capable of describing a wide range of frequency and density dependencies of selection and growth. Our analysis revealed the necessity of taking eco-evolutionary feedback into account in order to make accurate predictions. In particular, we found that the genetic composition of the population bulk may be a poor predictor for the genetic composition of the population front. Similarly, the measurements at the expansion front may be a poor predictor of the properties of the new population once it is fully established. These differences between the population bulk and expansion front make it also harder to predict the rates of expansions. Indeed, we found that the rates of expansions could be accelerating as the result of slow evolutionary changes at the expanding population edge and that species can spread faster in mixed waves than in isolation or by invading already established populations.  

Our main result is that colonization of new territories can proceed in two qualitatively different ways. The first possibility is a single mixed wave, where both alleles~(or species) move with the same velocity and their relative abundances reach a steady state in the reference frame comoving with the expansion. The second possibility is that one of the alleles~(or species) outruns the other. The faster allele is solely responsible for the colonization, while the slower allele invades the faster one from behind with a smaller velocity. As a result, there is a growing region occupied exclusively by the faster allele. The effect of wave splitting could be especially dramatic for species that have markedly smaller migration rates in the population bulk compared to population front~\cite{degiorgio:africa, korolev:sectors} because the invasion by the slower allele would be significantly slower than the range expansion. The existence of secondary genetic waves could lead to unexpected population and genetic dynamics, which cannot be described by the classic models of range expansions~\cite{kolmogorov:wave, fisher:wave, aronson:allee_wave, fife:allee_wave}. It would be interesting to know when the commonly observed changes behind the expansion front~\cite{phillips:toad_tropical} are caused by de novo adaptation to the new environment and when they are caused by the secondary genetic waves, which could reach and alter populations at the newly colonized territories many generations after the arrival of the population wave.  

In the context of cooperator-defector interactions, wave splitting and the increase in cooperation at the front could stabilize cooperation against defectors in populations that experience frequent disturbances followed by recolonizations. This mechanism of maintaining cooperation does not rely on reciprocity or multi-level selection, but is instead grounded in the density dependence of the evolutionary dynamics. More importantly, we have demonstrated that the coupling between ecological and evolutionary dynamics can have a profound effect on the fate of cooperation and should, therefore, be considered in both theoretical and experimental studies. \\

\section*{Acknowledgments}

\noindent
We would like to acknowledge useful discussions with Jeff Gore, Ivana Cvijovic, and Manoshi Datta. 

\bibliography{ws}

\begin{thebibliography}{10}
\providecommand{\url}[1]{\texttt{#1}}
\providecommand{\urlprefix}{URL }
\expandafter\ifx\csname urlstyle\endcsname\relax
  \providecommand{\doi}[1]{doi:\discretionary{}{}{}#1}\else
  \providecommand{\doi}{doi:\discretionary{}{}{}\begingroup
  \urlstyle{rm}\Url}\fi
\providecommand{\bibAnnoteFile}[1]{%
  \IfFileExists{#1}{\begin{quotation}\noindent\textsc{Key:} #1\\
  \textsc{Annotation:}\ \input{#1}\end{quotation}}{}}
\providecommand{\bibAnnote}[2]{%
  \begin{quotation}\noindent\textsc{Key:} #1\\
  \textsc{Annotation:}\ #2\end{quotation}}
\providecommand{\eprint}[2][]{\url{#2}}

\bibitem{gardner:cooperation_history}
Gardner A, Foster KR (2008) {The evolution and ecology of cooperation--history
  and concepts}.
\newblock In: Korb J, Heinze J, editors, Ecology of Social Evolution, Springer,
  Berlin. pp. 1--36.
\bibAnnoteFile{gardner:cooperation_history}

\bibitem{nowak:book}
Nowak M (2006) {Evolutionary dynamics: exploring the equations of life}.
\newblock Belknap Press.
\bibAnnoteFile{nowak:book}

\bibitem{smith:games}
{Maynard Smith} J (1982) {Evolution and the Theory of Games}.
\newblock Cambridge University Press, Cambridge UK.
\bibAnnoteFile{smith:games}

\bibitem{west:social}
West SA, Diggle SP, Buckling A, Gardner A, Griffin AS (2007) {The social lives
  of microbes}.
\newblock Annu Rev Ecol Evol Syst 38: 53--77.
\bibAnnoteFile{west:social}

\bibitem{holldobler:ants}
H{\"o}lldobler B, Wilson E (1990) The ants.
\newblock Belknap Press.
\bibAnnoteFile{holldobler:ants}

\bibitem{michod:multicellularity}
Michod RE, Roze D (2001) {Cooperation and conflict in the evolution of
  multicellularity}.
\newblock Heredity 86: 1--7.
\bibAnnoteFile{michod:multicellularity}

\bibitem{sachs:multicellularity}
Sachs JL (2008) {Resolving the first steps to multicellularity}.
\newblock Trends Evol Evol 23: 245--248.
\bibAnnoteFile{sachs:multicellularity}

\bibitem{hardin:tragedy}
Hardin G (1968) {The tragedy of the commons}.
\newblock Science 162: 1243--1248.
\bibAnnoteFile{hardin:tragedy}

\bibitem{crespi:cancer_evolution}
Crespi B, Summers K (2005) {Evolutionary biology of cancer}.
\newblock Trends Ecol Evol 20: 545--552.
\bibAnnoteFile{crespi:cancer_evolution}

\bibitem{merlo:cancer_evolution}
Merlo LMF, Pepper JW, Reid BJ, Maley CC (2006) {Cancer as an evolutionary and
  ecological process}.
\newblock Nat Rev Cancer 6: 924--935.
\bibAnnoteFile{merlo:cancer_evolution}

\bibitem{nowak:five_rules}
Nowak MA (2006) {Five rules for the evolution of cooperation.}
\newblock Science 314: 1560--1563.
\bibAnnoteFile{nowak:five_rules}

\bibitem{axelrod:cooperation}
Axelrod R, Hamilton WD (1981) {The evolution of cooperation}.
\newblock Science 211: 1390--1396.
\bibAnnoteFile{axelrod:cooperation}

\bibitem{nowak:generous_tft}
Nowak MA, Sigmund K (1992) {Tit for tat in heterogeneous populations}.
\newblock Nature 355: 250--253.
\bibAnnoteFile{nowak:generous_tft}

\bibitem{wilson:group_selection}
Wilson D (1975) A theory of group selection.
\newblock Proc Natl Acad Sci U S A 72: 143--146.
\bibAnnoteFile{wilson:group_selection}

\bibitem{hamilton:rule1}
Hamilton WD (1964) The genetical evolution of social behaviour. i.
\newblock J Theor Biol 7: 1--16.
\bibAnnoteFile{hamilton:rule1}

\bibitem{hamilton:rule2}
Hamilton WD (1964) The genetical evolution of social behaviour. ii.
\newblock J Theor Biol 7: 17--52.
\bibAnnoteFile{hamilton:rule2}

\bibitem{nowak:spatial}
Nowak MA, May RM (1992) {Evolutionary games and spatial chaos}.
\newblock Nature 359: 826--829.
\bibAnnoteFile{nowak:spatial}

\bibitem{roca:spatial_cooperation}
Roca C, Cuesta J, Sanchez A (2009) {Effect of spatial structure on the
  evolution of cooperation}.
\newblock Phys Rev E Stat Nonlin Soft Matter Phys 80: 046106.
\bibAnnoteFile{roca:spatial_cooperation}

\bibitem{nadell:EmergenceSpatialStructure}
Nadell CD, Foster KR, Xavier J (2010) {Emergence of spatial structure in cell
  groups and the evolution of cooperation}.
\newblock PLoS Comput Biol 6: e1000716.
\bibAnnoteFile{nadell:EmergenceSpatialStructure}

\bibitem{turner:rna}
Turner PE, Chao L (1999) {Prisoner's dilemma in an RNA virus.}
\newblock Nature 398: 441--443.
\bibAnnoteFile{turner:rna}

\bibitem{rainey:surface_film}
Rainey PB, Rainey K (2003) {Evolution of cooperation and conflict in
  experimental bacterial populations.}
\newblock Nature 425: 72--74.
\bibAnnoteFile{rainey:surface_film}

\bibitem{griffin:pathogenic}
Griffin AS, West SA, Buckling A (2004) {Cooperation and competition in
  pathogenic bacteria.}
\newblock Nature 430: 1024--1027.
\bibAnnoteFile{griffin:pathogenic}

\bibitem{hauert:prs_public_goods}
Hauert C, Holmes M, Doebeli M (2006) {Evolutionary games and population
  dynamics: maintenance of cooperation in public goods games}.
\newblock Proc Biol Sci 273: 2565--2571.
\bibAnnoteFile{hauert:prs_public_goods}

\bibitem{hauert:tpb_public_goods}
Hauert C, Wakano JY, Doebeli M (2008) {Ecological public goods games:
  Cooperation and bifurcation}.
\newblock Theor Popul Biol 73: 257--263.
\bibAnnoteFile{hauert:tpb_public_goods}

\bibitem{wakano:spatial_public_goods}
Wakano JY, Nowak MA, Hauert C (2009) {Spatial dynamics of ecological public
  goods.}
\newblock Proc Natl Acad Sci U S A 106: 7910--7914.
\bibAnnoteFile{wakano:spatial_public_goods}

\bibitem{shine:spatial_sorting}
Shine R, Brown GP, Phillips BL (2011) {An evolutionary process that assembles
  phenotypes through space rather than through time}.
\newblock Proc Natl Acad Sci U S A 108: 5708--5711.
\bibAnnoteFile{shine:spatial_sorting}

\bibitem{templeton:africa}
Templeton A (2002) {Out of Africa again and again}.
\newblock Nature 416: 45--51.
\bibAnnoteFile{templeton:africa}

\bibitem{parmesan:butterfly_climate}
Parmesan C, Ryrholm N, Stefanescu C, Hill JK, Thomas CD, et~al. (1999)
  {Poleward shifts in geographical ranges of butterfly species associated with
  regional warming}.
\newblock Nature 399: 579--583.
\bibAnnoteFile{parmesan:butterfly_climate}

\bibitem{pateman:butterfly_climate}
Pateman RM, Hill JK, Roy DB, Fox R, Thomas CD (2012) {Temperature-Dependent
  Alterations in Host Use Drive Rapid Range Expansion in a Butterfly}.
\newblock Science 336: 1028-1030.
\bibAnnoteFile{pateman:butterfly_climate}

\bibitem{gray:worm_expansion}
Gray ME, Sappington TW, Miller NJ, Moeser J, Bohn MO (2009) {Adaptation and
  Invasiveness of Western Corn Rootworm: Intensifying Research on a Worsening
  Pest}.
\newblock Annu Rev Entomol 54: 303--321.
\bibAnnoteFile{gray:worm_expansion}

\bibitem{phillips:toad_acceleration}
Phillips BL, Brown GP, Webb JK, Shine R (2006) {Invasion and the evolution of
  speed in toads}.
\newblock Nature 439: 803--803.
\bibAnnoteFile{phillips:toad_acceleration}

\bibitem{hitch:climate_birds}
Hitch AT, Leberg PL (2007) {Breeding Distributions of North American Bird
  Species Moving North as a Result of Climate Change}.
\newblock Conserv Biol 21: 534--539.
\bibAnnoteFile{hitch:climate_birds}

\bibitem{bronnenhuber:goby_invasion}
Bronnenhuber JE, Dufour BA, Higgs DM, Heath DD (2011) {Dispersal strategies,
  secondary range expansion and invasion genetics of the nonindigenous round
  goby, Neogobius melanostomus, in Great Lakes tributaries}.
\newblock Mol Ecol 20: 1845--1859.
\bibAnnoteFile{bronnenhuber:goby_invasion}

\bibitem{excoffier:review}
Excoffier L, Foll M, Petit RJ (2009) {Genetic consequences of range
  expansions.}
\newblock Annu Rev Ecol Evol Syst 40: 481--501.
\bibAnnoteFile{excoffier:review}

\bibitem{korolev:review}
Korolev KS, Avlund M, Hallatschek O, Nelson DR (2010) {Genetic demixing and
  evolution in linear stepping stone models}.
\newblock Rev Mod Phys 82: 1691--1718.
\bibAnnoteFile{korolev:review}

\bibitem{vlad:hydrodynamics}
Vlad MO, Cavalli-Sforza LL, Ross J (2004) {Enhanced (hydrodynamic) transport
  induced by population growth in reaction-diffusion systems with application
  to population genetics}.
\newblock Proc Natl Acad Sci U S A 101: 10249--10253.
\bibAnnoteFile{vlad:hydrodynamics}

\bibitem{korolev:mutualism}
Korolev KS, Nelson D (2011) {Competition and Cooperation in One-Dimensional
  Stepping-Stone Models}.
\newblock Phys Rev Lett 107: 0881031--0881035.
\bibAnnoteFile{korolev:mutualism}

\bibitem{mcinerny:climate_shift}
McInerny GJ, Turner JRG, Wong HY, Travis JMJ, Benton TG (2009) {How range
  shifts induced by climate change affect neutral evolution}.
\newblock Proc Biol Sci 276: 1527--1534.
\bibAnnoteFile{mcinerny:climate_shift}

\bibitem{roques:allee_diversity}
Roques L, Garnier J, Hamel F, Klein EK (2012) {Allee effect promotes diversity
  in traveling waves of colonization}.
\newblock Proc Natl Acad Sci U S A 109: 8828--8833.
\bibAnnoteFile{roques:allee_diversity}

\bibitem{arenas:diversity}
Arenas M, Ray N, Currat M, Excoffier L (2011) {Consequences of Range
  Contractions and Range Shifts on Molecular Diversity}.
\newblock Mol Biol Evol 29: 207--218.
\bibAnnoteFile{arenas:diversity}

\bibitem{hallatschek:1dwave}
Hallatschek O, Nelson D (2008) {Gene surfing in expanding populations}.
\newblock Theor Popul Biol 73: 158--170.
\bibAnnoteFile{hallatschek:1dwave}

\bibitem{hallatschek:life_front}
Hallatschek O, Nelson DR (2010) {Life at the front of an expanding population}.
\newblock Evolution 64: 193--206.
\bibAnnoteFile{hallatschek:life_front}

\bibitem{travis:deleterious_surfing}
Travis JMJ, Munkemuller T, Burton OJ, Best A, Dytham C, et~al. (2007)
  {Deleterious Mutations Can Surf to High Densities on the Wave Front of an
  Expanding Population}.
\newblock Mol Biol Evol 24: 2334--2343.
\bibAnnoteFile{travis:deleterious_surfing}

\bibitem{murray:mathematical_biology}
Murray JD (2003) {Mathematical Biology}.
\newblock Springer.
\bibAnnoteFile{murray:mathematical_biology}

\bibitem{douglas:assembly_wave}
Douglas JF, Efimenko K, Fischer DA, Phelan FR, Genzer J (2007) Propagating
  waves of self-assembly in organosilane monolayers.
\newblock Proc Natl Acad Sci U S A 104: 10324--10329.
\bibAnnoteFile{douglas:assembly_wave}

\bibitem{sella:dynamic_persistence}
Sella G, Lachmann M (2000) {On the Dynamic Persistence of Cooperation: How
  Lower Individual Fitness Induces Higher Survivability}.
\newblock J Theor Biol 206: 465--485.
\bibAnnoteFile{sella:dynamic_persistence}

\bibitem{fisher:wave}
Fisher RA (1937) {The wave of advance of advantageous genes}.
\newblock Ann Eugen 7: 355--369.
\bibAnnoteFile{fisher:wave}

\bibitem{kolmogorov:wave}
Kolmogorov AN, Petrovsky N, Piscounov NS (1937) {A study of the equation of
  diffusion with increase in the quantity of matter, and its application to a
  biological problem}.
\newblock Moscow University Bulletin of Mathematics 1: 1.
\bibAnnoteFile{kolmogorov:wave}

\bibitem{allee:effect}
Allee W, Bowen E (1932) Studies in animal aggregations: mass protection against
  colloidal silver among goldfishes.
\newblock J Exp Zool 61: 185--207.
\bibAnnoteFile{allee:effect}

\bibitem{courchamp:allee_review}
Courchamp F, Clutton-Brock T, Grenfell B (1999) {Inverse density dependence and
  the Allee effect}.
\newblock Trends Ecol Evol 14: 405--410.
\bibAnnoteFile{courchamp:allee_review}

\bibitem{berec:multiple_allee}
Berec L, Angulo E, Courchamp F (2007) {Multiple Allee effects and population
  management}.
\newblock Trends Ecol Evol 22: 185--191.
\bibAnnoteFile{berec:multiple_allee}

\bibitem{dai:warnings}
Dai L, Vorselen D, Korolev KS, Gore J (2012) {Generic indicators for loss of
  resilience before a tipping point leading to population collapse.}
\newblock Science 336: 1175--1177.
\bibAnnoteFile{dai:warnings}

\bibitem{courchamp:dog}
Courchamp F, Macdonald DW (2001) {Crucial importance of pack size in the
  African wild dog Lycaon pictus}.
\newblock Animal Conservation 4: 169--174.
\bibAnnoteFile{courchamp:dog}

\bibitem{cluttonbrock:mongoose}
Clutton~Brock TH, Gaynor D, McIlrath GM, Maccoll A, Kansky R, et~al. (1999)
  {Predation, group size and mortality in a cooperative mongoose, Suricata
  suricatta}.
\newblock J Anim Ecol 68: 672--683.
\bibAnnoteFile{cluttonbrock:mongoose}

\bibitem{mooring:sheep}
Mooring MS, Fitzpatrick TA, Nishihira TT, Reisig DD (2004) {Vigilance,
  predation risk, and the Allee effect in desert bighorn sheep}.
\newblock J Wildl Manage 68: 519--532.
\bibAnnoteFile{mooring:sheep}

\bibitem{fife:allee_wave}
Fife PC, McLeod JB (1977) {The approach of solutions of nonlinear diffusion
  equations to travelling front solutions}.
\newblock Arch Ration Mech Anal 65: 335--361.
\bibAnnoteFile{fife:allee_wave}

\bibitem{aronson:allee_wave}
Aronson DG, Weinberger HG (1975) Nonlinear diffusion in population genetics,
  combustion and nerve propagation Lectures Notes Math, volume 446.
\newblock Springer, New York, 5–49 pp.
\bibAnnoteFile{aronson:allee_wave}

\bibitem{frey:review}
Frey E (2010) {Evolutionary game theory: Theoretical concepts and applications
  to microbial communities}.
\newblock Physica A 389: 4265--4298.
\bibAnnoteFile{frey:review}

\bibitem{gore:snow_drift}
Gore J, Youk H, van Oudenaarden A (2009) {Snowdrift game dynamics and
  facultative cheating in yeast}.
\newblock Nature 459: 253--256.
\bibAnnoteFile{gore:snow_drift}

\bibitem{macLean:high_k}
MacLean RC, Fuentes-Hernandez A, Greig D, Hurst LD, Gudelj I (2010) {A Mixture
  of ``Cheats'' and ``Co-Operators'' Can Enable Maximal Group Benefit}.
\newblock PLoS Biol 8: e1000486.
\bibAnnoteFile{macLean:high_k}

\bibitem{sanchez:yeast_dynamics}
Sanchez A, Gore J (2013) Feedback between population and evolutionary dynamics
  determines the fate of social microbial populations Available:
  http://arxiv.org/abs/1301.2791. Accessed 14 February 2013.
\bibAnnoteFile{sanchez:yeast_dynamics}

\bibitem{press:numerical_recipes}
Press WH, Teukolsky SA, Vetterling WT, Flannery BP (2007) Numerical Recipes:
  The Art of Scientific Computing.
\newblock Cambridge University Press, Cambridge.
\bibAnnoteFile{press:numerical_recipes}

\bibitem{hilker:pathogens}
Hilker FM, Lewis MA, Seno H, Langlais M, Malchow H (2005) Pathogens can slow
  down or reverse invasion fronts of their hosts.
\newblock Biol Invasions 7: 817--832.
\bibAnnoteFile{hilker:pathogens}

\bibitem{titchmarsh:second_order_ode}
Titchmarsh E (1946) Eigenfunction expansions associated with second-order
  differential equations.
\newblock Clarendon Press, Oxford.
\bibAnnoteFile{titchmarsh:second_order_ode}

\bibitem{merzbacher:qm}
Merzbacher E (1997) Quantum Mechanics.
\newblock Wiley.
\bibAnnoteFile{merzbacher:qm}

\bibitem{hallatschek:fisher_wave}
Hallatschek O, Korolev KS (2009) {Fisher Waves in the Strong Noise Limit}.
\newblock Phys Rev Lett 103: 108103--108106.
\bibAnnoteFile{hallatschek:fisher_wave}

\bibitem{gillespie:textbook}
Gillespie J (2004) Population genetics: a concise guide.
\newblock Johns Hopkins University Press.
\bibAnnoteFile{gillespie:textbook}

\bibitem{thomas:acceleration}
Thomas CD, Bodsworth EJ, Wilson RJ, Simmons AD, Davies ZG, et~al. (2001)
  {Ecological and evolutionary processes at expanding range margins}.
\newblock Nature 411: 577--581.
\bibAnnoteFile{thomas:acceleration}

\bibitem{phillips:toad_tropical}
Phillips BL, Brown GP, Greenlees M, Webb JK, Shine R (2007) {Rapid expansion of
  the cane toad (Bufo marinus) invasion front in tropical Australia}.
\newblock Austral Ecology 32: 169--176.
\bibAnnoteFile{phillips:toad_tropical}

\bibitem{kokko:ecoevo}
Kokko H, L{\'o}pez-Sepulcre A (2007) The ecogenetic link between demography and
  evolution: can we bridge the gap between theory and data?
\newblock Ecol Lett 10: 773--782.
\bibAnnoteFile{kokko:ecoevo}

\bibitem{degiorgio:africa}
DeGiorgio M, Jakobsson M, Rosenberg N (2009) Explaining worldwide patterns of
  human genetic variation using a coalescent-based serial founder model of
  migration outward from africa.
\newblock Proc Natl Acad Sci U S A 106: 16057--16062.
\bibAnnoteFile{degiorgio:africa}

\bibitem{korolev:sectors}
Korolev KS, M\"uller MJI, Karahan N, Murray O Andrew W~Hallatschek, Nelson DR
  (2012) Selective sweeps in growing microbial colonies.
\newblock Phys Biol 9: 026008--026023.
\bibAnnoteFile{korolev:sectors}

\bibitem{saarloos:review}
Van~Saarloos W (2003) {Front propagation into unstable states}.
\newblock Phys Rep 386: 29--222.
\bibAnnoteFile{saarloos:review}

\end{thebibliography}

\clearpage

\section*{Text S1}

\renewcommand{\theequation}{S\arabic{equation}}
\setcounter{equation}{0}
\noindent In this Supporting Text, we first expand our discussion of the boundary conditions for equation~(16). We then outline the derivation of the exact solution, given by equation~(18), for the transition from non-splitting to splitting behavior in a special case of equation~(12). Finally, we show that both effective advection and effective growth can allow defectors to expand faster in mixed waves than they can do by invading cooperators.\\


\noindent \textbf{Boundary conditions for equation~(16)}\\

\noindent Here, we show that~$\psi(\pm\infty)=0$ are the proper boundary conditions for equation~(16). Since the boundary conditions for~$f(\zeta)$ are~$f(-\infty)=1$ and~$f(+\infty)=0$, it is sufficient to show that~$u(-\infty)>0$ and~$u(+\infty)\le0$; see equation~(15). The former condition follows from the fact that~$v_{\rm{c}}>0$ and~$v_{\rm{a}}(-\infty)=0$. The latter condition follows from the fact that~$v_{\rm{a}}(+\infty)\ge v_{\rm{c}}$, which we prove below. 

The density profile at large~$\zeta$ decays exponentially as~$e^{-\gamma\zeta}$; moreover,~$\gamma\ge\sqrt{G_{c}(0,0)/D}$ for a weak Allee effect~\cite{saarloos:review}. We then substitute this scaling form in the equation~(9) for~$c(t,x)$, change to the comoving reference frame, and require that the wave profile does not depend on~$\tau$. This calculation yields

\begin{equation}
	v_{\rm{c}} = D\gamma + \frac{G_{c}(0,0)}{\gamma}.
	\label{eq:gamma_v}
\end{equation}

\noindent This expression for~$v_{\rm{c}}$ can now be compared to the advection velocity~$v_{\rm{a}}=-2D\partial\ln[c(t,x)]/\partial x = 2D\gamma$. For a strong Allee effect,~$G_{c}(0,0)$ is negative and, therefore,~$v_{\rm{a}}>v_{\rm{c}}$. For a weak Allee effect,~$v_{\rm{a}}-v_{\rm{c}}=D\gamma-\frac{G_{c}(0,0)}{\gamma}\ge0$, where the last inequality follows from~$\gamma\ge\sqrt{G_{c}(0,0)/D}$; see~\cite{saarloos:review}.\\

\noindent \textbf{Derivation of equation~(18)} \\

\noindent
Here, we derive equation~(18). Since, for~$G_{f}(c,f)$ decreasing with~$f$, wave splitting is affected only by the linearization of~$G_{f}(c,f)f$ for small~$f$, we are free to replace~$G_{f}(c,f)f$ with any other function that has identical behavior for~$f\to0$. It is particularly useful to approximate~$G_{f}(c,f)f$ with a piece-wise linear function because this allows for an exact solution for~$f(t,x)$ everywhere in space, which captures the qualitative dependence of the solution on the parameters of the model. In contrast, eigenfunctions of~$\mathcal{L}$ only describe the behavior of~$f(t,x)$ close to the front and only up to a normalization factor. To this purpose, we redefine~$G(c,f)f$ as

\begin{equation}
	G_{f}(c,f)f=g_{f}(c)f^{*}(c)\min\{f,f^{*}(c)-f\}.
	\label{eq:gf_linear}
\end{equation}

\noindent Note that this redefinition does not have a fixed point at~$f=1$, which is appropriate as long as we are not interested in cooperators invading defectors. 

We now turn to equation~(12), which can be further simplified by the change of variables from~$\zeta$ to~$\varphi=c/K$ using equation~(4). The result reads

\begin{equation}
	\varphi^{2}(1-\varphi)^{2}f'' + 2\varphi(1-\varphi)\left(1-2\varphi+\frac{c^{*}}{K}\right) + \frac{2}{g_{c}K^{2}}[G_{f}(K\varphi,f)-\lambda]f=0,
	\label{eq:change_phi}
\end{equation}

\noindent where primes now denote derivatives with respect to~$\varphi$. We further assume that~$c^{*}=0$ and obtain that

\begin{equation}
	[\varphi^{2}(1-\varphi)^{2}f']'=-\frac{2}{g_{c}K^{2}}G_{f}(K\varphi,f)f,	
	\label{eq:c_star_zero}
\end{equation}

\noindent where we also set~$\lambda=0$ because~$\lambda$ vanishes at the splitting transition. We now use equation~(\ref{eq:gf_linear}) and the assumption that~$f^{*}(\varphi)=f^{*}$ for~$\varphi>\bar{\varphi}\equiv\bar{c}/K$ and~$f(\varphi)=0$ for~$\varphi<\bar{\varphi}$, i.e. defectors are not viable at very low densities. The right hand side of equation~(\ref{eq:c_star_zero}) is then proportional to~$f$ for~$f<f^{*}/2$ and to~$f^{*}-f$ for~$f>f^{*}/2$; therefore, further simplifications occur upon integrating both sides of equation~(\ref{eq:c_star_zero}) over~$\varphi$ and defining a new variable~$F=\int fd\varphi$ for~$f<f^{*}/2$ and~$E=\int (f^{*}-f)d\varphi$ for~$f>f^{*}/2$. For example, for the region where~$f<f^{*}/2$ and~$\varphi>\bar{\varphi}$, we obtain

\begin{equation}
	\varphi^{2}(1-\varphi)^{2}F'' = -\frac{2g_{f}f^{*}}{g_{c}K^{2}}F
	\label{eq:f_int}
\end{equation}

\noindent The general solution of this equation is given by

\begin{gather}
\begin{aligned}
	& F=C_{1}\varphi^{\alpha}(1-\varphi)^{1-\alpha}+C_{2}\varphi^{1-\alpha}(1-\varphi)^{\alpha},\;\; \mbox{with} \\
	& \alpha=\frac{1}{2}\left(1+\sqrt{1-\frac{8g_{f}f^{*}}{g_{c}K^{2}}}\right).
\end{aligned}
	\label{eq:general_solution}
\end{gather}

\noindent Similarly, for~$f>f^{*}/2$, we find

\begin{gather}
\begin{aligned}
	& E=C_{3}\varphi^{\beta}(1-\varphi)^{1-\beta}+C_{4}\varphi^{1-\beta}(1-\varphi)^{\beta},\;\; \mbox{with} \\
	& \beta=\frac{1}{2}\left(1+\sqrt{1+\frac{8g_{f}f^{*}}{g_{c}K^{2}}}\right).
\end{aligned}
	\label{eq:general_solution_large_f}
\end{gather}

Then, the solution for~$f(\varphi)$ is given by~$E'$ from equation~(\ref{eq:general_solution_large_f}) for~$f>f^{*}/2$ and by~$F'$ from equation~(\ref{eq:general_solution}) for~$f<f^{*}/2$ and~$\varphi>\bar{\varphi}$. The four constants~$C_{1}$, $C_{2}$, $C_{3}$, and~$C_{4}$ and the matching point~$\varphi_{1/2}$ are determined by the boundary conditions~$f(-\infty)=f^{*}$ and~$f(\bar{c}/K)=0$ and the matching conditions at~$\varphi_{1/2}$, where~$f(\varphi_{1/2}-0)=f(\varphi_{1/2}+0)=1/2$ and~$f'(\varphi_{1/2}-0)=f'(\varphi_{1/2}+0)$. Upon using four of the five conditions, one can easily express the four constants in terms of~$\varphi_{1/2}$ because the conditions are linear with respect to~$C_{1}$, $C_{2}$, $C_{3}$, and~$C_{4}$. The remaining condition determines~$\varphi_{1/2}$ and can also be solved exactly:

\begin{gather}
\begin{aligned}
	& \varphi_{1/2}=\frac{\gamma\bar{\varphi}}{1-(1-\gamma)\bar{\varphi}},\;\;\mbox{where} \\
	& \gamma=\left[\frac{(\beta-\alpha)(\alpha-\bar{\varphi})}{(\beta+\alpha-1)(1-\alpha-\bar{\varphi})}\right]^{\frac{1}{2\alpha-1}}.
\end{aligned}
	\label{eq:phi_half}
\end{gather}

\noindent We then obtain the condition for splitting, stated in equation~(18), by requiring that~$\gamma$ is real, and~$\varphi_{1/2}>\bar{\varphi}$. Note that,~$\varphi_{1/2}\to1$ as the splitting state is approached, say, by increasing~$\bar{c}$. Since~$\varphi\to1$ corresponds to~$\zeta\to-\infty$, we see that the wave of defectors lags more and more behind the wave of cooperators. As we show in the main text, this is a general result, and the front is populated mostly by the cooperators close to the splitting transition. As a consequence, the expansion velocity at the splitting transition is given by equation~(3).\\

\noindent \textbf{Increased spreading through effective advection and growth}\\

\noindent
Here, we demonstrate the effects of effective advection and growth on the ability of defectors to keep up with cooperators. When ecological dynamics are not affected by the evolutionary dynamics, the ability of defectors to follow cooperators in a mixed wave is mathematically equivalent to the ability of a Fisher wave to follow a moving boundary between good and bad conditions. In particular, the assumption that~$f=0$ for~$c<\bar{c}$ makes this moving boundary absorbing. One can easily show that classic Fisher waves can only follow an absorbing boundary that moves with a velocity smaller than the Fisher velocity. Here, we show that even boundaries moving with higher velocities can be followed provided there is an effective advection or growth. This analysis illustrates the role of the term~$2D\frac{\partial \ln(c)}{\partial x}\frac{\partial f}{\partial x}$ in equation~(10) and suggests that both effective advection and growth could be useful management strategies to maintain genetic diversity during range expansions or to help species shift in response to a rapid climate change.

To avoid confusion with the earlier discussion, we introduce new notation. Let~$v$ be the velocity of the absorbing barrier, and~$g$ is the per capita growth rate. The Fisher velocity is then~$2\sqrt{gD}$, where~$D$ is the effective diffusion constant. When~$v>2\sqrt{gD}$, the population cannot keep up with the barrier without external interventions. One possible intervention is to increase~$g$ by a factor~$1+\delta$, say by supplying extra food, in a region of length~$a$ behind the barrier, which is analogous to the effective growth discussed earlier. Another possibility, analogous to effective advection, is to move the population with velocity~$u$ in a region of length~$a$ behind the front, say by capturing and transporting individuals within the population. The analysis of both of these scenarios follows the same procedure as the derivation of~(18), but is much simpler because the resulting linearized equations have constant coefficients; therefore they can be easily solved by the standard methods. Below we just summarize the results. 

For effective growth, we find that barriers moving with any velocity can be followed, but the region of effective growth should be sufficiently large:

\begin{equation}
	a>\frac{2D}{\sqrt{4gD(1+\delta)-v^{2}}}\left[\pi-\arctan\left(\frac{\sqrt{4gD(1+\delta)-v^{2}}}{\sqrt{v^{2}-4gD}}\right)\right].
	\label{eq:a_growth}
\end{equation}

\noindent Note that the minimal size is finite when~$v=2\sqrt{gD}$, and that it decreases only as~$\delta^{-1/2}$ for large~$\delta$. This suggests that there is an optimal size~$a$ that minimizes~$\delta a$, which could be interpreted as the total cost of the intervention.

For effective advection, we also find that barriers moving with any velocity can be followed provided~$2\sqrt{gD}+u>v$ and the region of advection is sufficiently large:

\begin{equation}
	a>\frac{2D}{\sqrt{4gD-(v-u)^{2}}}\left[\pi-\arctan\left(\frac{\sqrt{4gD-(v-u)^{2}}}{u+\sqrt{v^{2}-4gD}}\right)\right].
	\label{eq:a_advection}
\end{equation}

\noindent Similar to the previous case, minimal~$a$ is finite when~$v=2\sqrt{gD}$. More importantly, the minimal size is the smallest when~$u$ and~$v$ are about equal. Indeed, when~$v\gg u$, advection has a small effect on the dynamics, while, when~$u\gg v$, the point where the region of advection ends becomes an effectively absorbing boundary because organisms entering advection region are quickly moved towards the front of the wave.\\

\end{document}